\documentclass[fleqn,10pt]{wlscirep}
\usepackage[utf8]{inputenc}
\usepackage[T1]{fontenc}
\usepackage{xr} 
\title{Single photon zeptosecond interferometry}

\author[1]{Geoffrey R. Harrison} \author[1]{Tobias Saule}
\author[1]{R. Esteban Goetz}  \author[1]{George N. Gibson}
\author[2]{Camilo Granados} 
\author[3,4,5]{Bikash K. Das}
\author[3,4,5]{Marcelo F. Ciappina}
\author[1]{Anh-Thu Le} 
\author[1,*]{Carlos A. Trallero-Herrero}

\affil[1]{Department of Physics, University of Connecticut, Storrs, Connecticut 06268, USA}
\affil[2]{Eastern Institute of Technology, Ningbo 315200, China}
\affil[3]{Department of Physics, Technion -- Israel Institute of Technology, Haifa 3200003, Israel}
\affil[4]{Department of Physics, Guangdong Technion -- Israel Institute of Technology, Shantou 515063, Guangdong, China}
\affil[5]{Guangdong Provincial Key Laboratory of Materials and Technologies for Energy Conversion, Guangdong Technion -- Israel Institute of Technology, Shantou 515063, Guangdong, China}

\affil[*]{carlos.trallero@uconn.edu}

\begin{abstract}
We demonstrate the generation of a train of attosecond XUV pulses that are in a superposition of wavefront states. 
Such superposition yields a high precision, self-referencing, common path XUV interferometer setup to produce pairs of spatially separated and independently controllable XUV pulses that are locked in phase and time with a temporal jitter of 3.5~zs (zs = zeptoseconds = $10^{-21}$s ). In our approach, we can independently control the relative phase/delay of the two optical beams with a resolution of 52~zs. Since the jitter is on the order of the Compton time scale $\tau_C=\lambda_C/m_ec^2$, we explore the level of correlation between the non-local photons by comparing different spatial mode superpositions. Further, thanks to the stability of the interferometer we can retrieve the interference pattern through photon counting. Through post-selection of different particle events we can analyze one, two or more photon events. We argue that this zeptosecond level of temporal precision will open the door for new dynamical QED tests at lower intensities while photon counting experiments can also have an impact on the emerging field of quantum light in strong fields. We also discuss the potential impact on other areas, such as time-dependent QED, imaging, measurements of non-locality, and molecular quantum tomography.
\end{abstract}
\begin{document}

\flushbottom
\maketitle
\thispagestyle{empty}

\section*{Introduction}
Over the last few decades, the timescales for observing new ultrafast phenomena have been ever-decreasing, mainly driven by new laser technologies, from Q-switching to attosecond pulses \cite{Paul2001}, the latter made possible through the process of higher-order harmonic generation (HHG) \cite{lewenstein1994theory,l1993high,krause1992high,orfanos2019attosecond}. While the 21st century saw the arrival of measurements at the atomic unit time scale of 24~as, a new temporal frontier is the Compton time scale $\tau_C=\lambda_C/c=8.2~zs$ (zeptosecond, zs=$10^{-21}$s), dictated by the Heisenberg uncertainty principle, with  $\lambda_C=h/m_ec$ the Compton wavelength for an electron at rest. Access to this new time frontier would allow, for example, new tests of quantum electrodynamics (QED) processes such as radiation reactions \cite{Ritus1985,DiPiazza2012,Wistisen2018,Poder2018} in a completely new regime without the need for extreme intensity.

More recently, there has been interest in bridging the gap between strong field and quantum optics \cite{gorlach2020quantum,gorlachHighharmonicGenerationDriven2023,moiseyevConditionsAnalogQED2023,tzurGenerationSqueezedHighorder2024,de-la-penaQuantumElectrodynamicsHighHarmonic2025,tzurMeasuringControllingBirth2025a,sennaryAttosecondQuantumUncertainty2025}. This renewed interest is partly driven by the potential to enhance precision measurements by reducing the photon wavelength. Another very promising direction is the production of entangled, squeezed, or other quantum states of light at extreme fluence, such as those encountered in TW ultrafast lasers.

In this work, we present a high precision, self-referencing common path XUV interferometer setup, akin to Young's double slit, with two spatially separated and independently controllable XUV pulses that are locked in phase and time.
The XUV pulses are generated through HHG and form an attosecond pulse train.
The technique allows for temporal (phase) control of 52~zs (0.86~mrad) of 113~nm light pulses, as well as the ability to measure temporal events with a minimum Allan deviation of 3.5~zs (0.08~mrad at 88nm).

At the core, it is self-referencing that allows for high temporal stability over extended periods of time. Our reported zs stability opens a new path for temporal tests of QED, while the long term stability allows for photon counting statistics measured interferometrically, presenting a promising avenue towards realizing quantum optics in the XUV.

\section*{Results}
\begin{figure}[t]%
\centering
\includegraphics[width=0.9\textwidth]{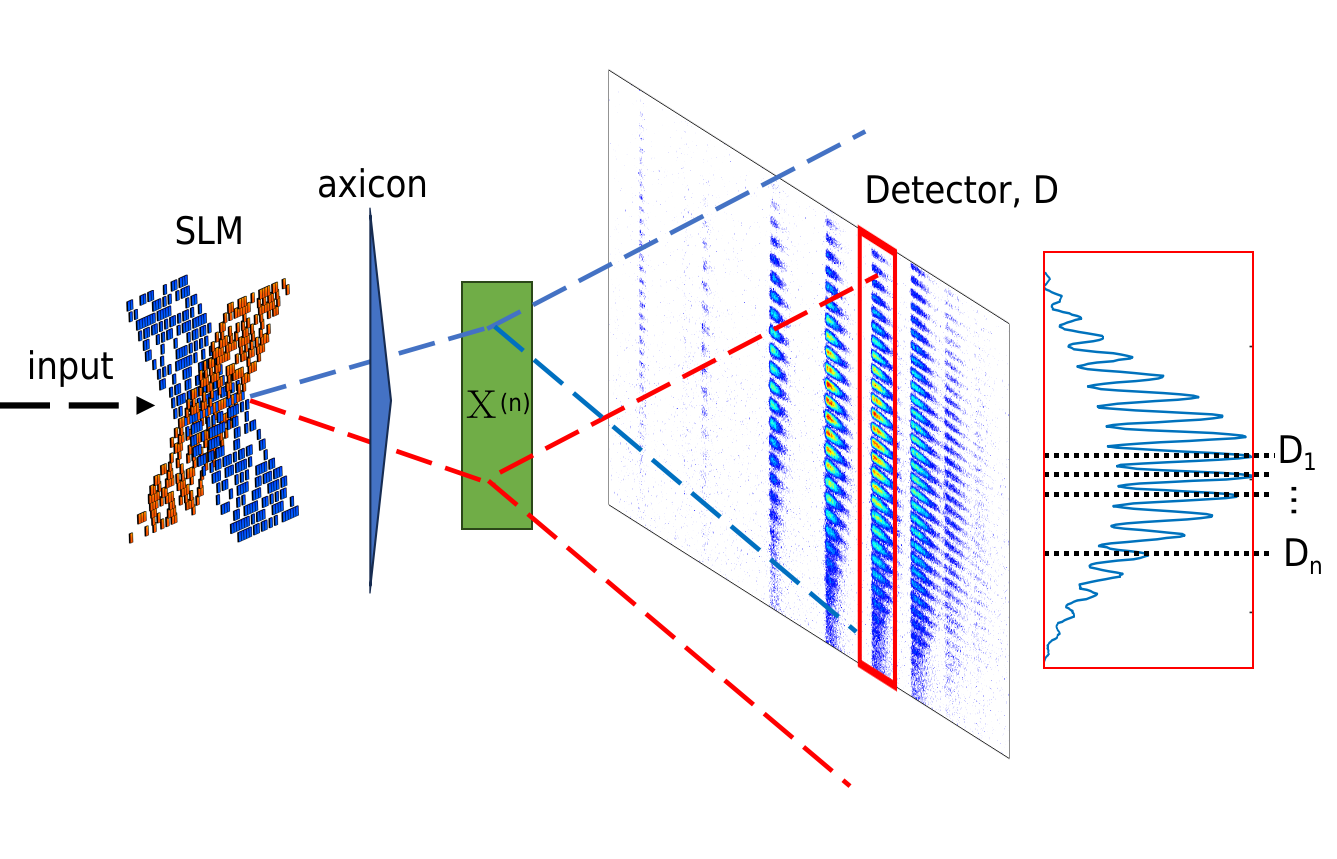}
\caption{
Sketch of the experimental setup. A spatial light modulator (SLM) is used to separate the incoming beam into multiple (here two) phase locked beams. They are used to generate spatially separated but phase locked harmonics which interfere in the far field and are detected with an XUV spectrometer. Note that each photon is in a superposition of two wavefronts after interacting with the SLM. $\chi^{(n)}$ represents an n-order nonlinearity.
} \label{fig:HOM_fig}
\end{figure}

Through HHG, an attosecond pulse train in the XUV ranging from the 7th to the 23rd harmonic of 800~nm light is generated. This fundamental is imprinted with two intertwined phase masks that result in two foci after focusing optics \cite{tross_interferometer,Tross2017,harrison2022increased}. 
In this work we examine three types of optical modes for attosecond interferometry; Gaussian, Bessel-Gaussian (BG) , and Laguerre-Gaussian (LG) modes. BG beams have the advantage of a vanishing far field which results in non-local HHG photons traveling through space with no other field present, making them ideally suited for field-free transient absorption spectroscopy or photoelectron spectroscopy. We call this approach interferometric transient absorption spectroscopy (ITAS).
A complete experimental setup is shown in the supplemental and is thoroughly described in the methods and reference \cite{harrison2022increased}, the latter also describing a method to increase the phase precision of a spatial light modulator (SLM) used in the experiments presented.
Here in short: two attosecond pulse trains interfere while spectrally separated in the far field.
In each harmonic phase differences between the wavefronts can be measured from movements of the interference fringes.
Samples can be placed into the arms of the interferometer and have their dipole interaction with the XUV light measured.
In this picture, each fringe can be interpreted as a separate detector, $D_n$, as in figure \ref{fig:HOM_fig}, and the phase measurement is made precise by taking information from all fringes together, as described in the supplementary section \ref{sect:phase_evaluation}

Initial testing with LG modes has shown that the two foci created with intertwined masks are formed from IR photons in a superposition of two wavefronts.
These experiments use the SLM to impart both $l=+1$ and $l=-1$ OAM onto the light which interferes in the focus creating two intense foci \cite{tross2019high}.
For LG with superimposed OAM,
\begin{equation}
E_{total}(r,\theta) = E_{1} r \mathrm{e}^{-\frac{r^2}{w_{0}^2}}\left[\mathrm{e}^{i\theta}+\mathrm{e}^{-i\theta}\right].
\label{eq:OAM}
\end{equation}
Focusing a field with distribution given by Eq. \ref{eq:OAM} yields two lobes at the focus with a net $l=0$.
A theoretical derivation is shown in the Supplemental materials.
This derivation is supported by the fact that the harmonics generated here do not exhibit the characteristic donut mode seen in beams with non-zero $l$ \cite{Gariepy_2014,tross2019high}.
Since the XUV photons have $l=0$ and OAM is conserved in HHG\cite{Kong2017, tross2019high}, the fundamental beams must also have net zero OAM. 
Furthermore, for this to be true of odd order harmonics, each photon of the fundamental must have net $l=0$, i.e. the SLM is putting photons into a superposition of $l=\pm 1$ states. 
As the BG foci used for ITAS are also generated with the SLM through similar mixed masks they will also be formed from photons in a superposition of states which should result in indistinguishable XUV photons, enhancing the stability of the interference fringes.

\begin{figure}[ht] 
\centering
\includegraphics[width=\textwidth]{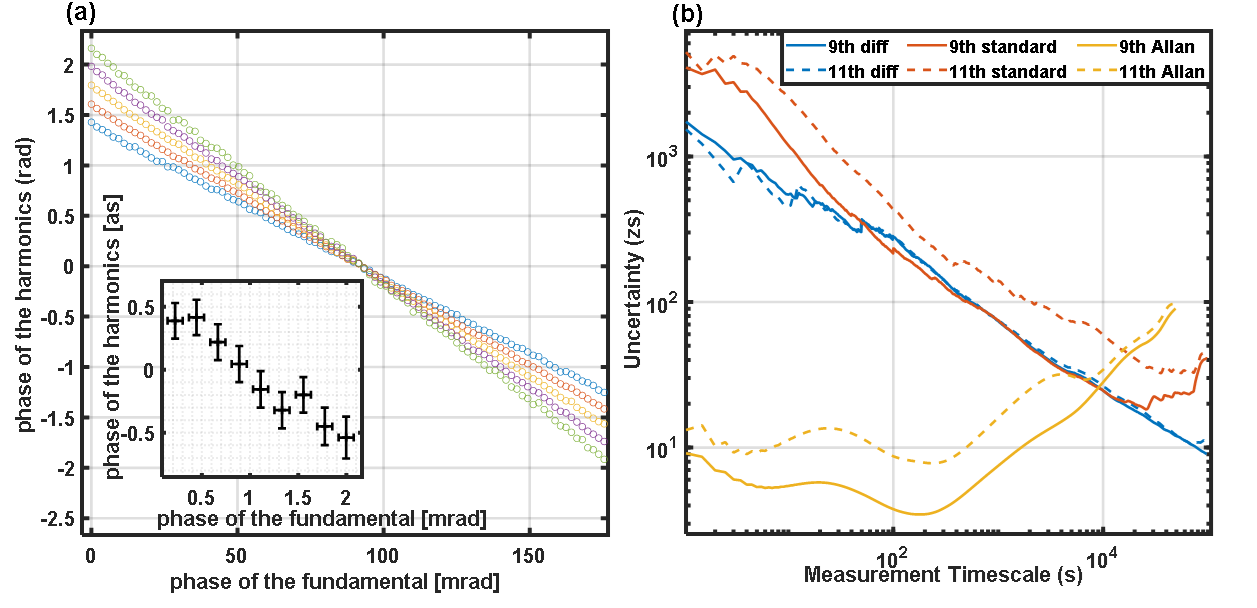}
\caption{(a) The phase evolution of harmonics 15-21, generated in Ar, as a function of fundamental phase difference. These harmonics show resolutions of around 75zs. Inset, a finer scan of the phase evolution in time of the 7th harmonic (113~nm) showing a resolution of 52~zs (0.86~mrad).
Here the resolution is close to the intended step size of the scan (84~zs) and is shown as the horizontal error bars. 
The vertical error bars show the precision achieved at each step.
(b) The error when characterizing the phase of a single mask as a function of the measurement time. The error is measured via three quantities.
First the standard errors $\sigma_s$ in red, which reaches a minimum value of 18~zs for the 9th harmonic. 
Then the differential errors $\sigma_D$ in blue, which reaches a minimum value of 9~zs after 26 hours. 
Finally, in yellow are the Allan deviations $\sigma_A$ for the two harmonics. The 9th reaches a minimum value of 3.5~zs after approximately 20 seconds. 
} 
\label{fig:stability}
\end{figure}

Figure \ref{fig:stability} shows that the stability of this interferometer reaches the single-digit zeptoseconds which enables single-photon XUV interferometry where the interferogram is recreated over several days of uninterrupted data acquisition or millions of laser shots with our 1~kHz source.
We argue that this capability will allow for future quantum-optics-like attosecond experiments.
Because of the fundamental nature of such experiments, we start by reminding the reader that a spectrometer provides a measurement of the time-average of the Fourier transform of the Poynting vector at point $\mathbf{r}_d$ on the detector plane and frequency $\omega$: 

\begin{equation}
\label{eq:poyntingS}
\langle \mathbf{S}_D\rangle(\mathbf{r}_d, \omega) = \Big\lvert \sum_{\alpha} E^{(\alpha)}_z(\mathbf{r}_d,\omega)\Big\rvert ^2 \, \text{\textbf{e}}_r,
\end{equation}
This, in turn, gives the well-known interferometric equation for two sources,
$\langle \mathbf{S}_D\rangle(\mathbf{r}_d, \omega) \!\!=\!\! I^{(1)}(\mathbf{r}_d,\omega) + I^{(2)}(\mathbf{r}_d,\omega) 
+I_{\text{int}}(\mathbf{r}_d,\omega)$, where, for this setup, $I_{\text{int}}(\mathbf{r}_d,\omega)$ is given by,
\begin{equation}
 I_{\text{int}}(\mathbf{r}_d,\omega) \propto  \big\lvert d^{(1)}_z(\omega)\big\rvert \,  \big\lvert d^{(2)}_z(\omega)\big\rvert\,
\cos\big[\phi^{(1)}_{dip}(\omega) -\phi^{(2)}_{dip}(\omega)  + \Delta\theta_{\text{opt}}(\mathbf{r}_d,\omega)\big],
\label{eq:Iint}
\end{equation}
with 
\begin{equation}
\phi^{(\alpha)}_{dip}(\omega)\equiv\arg\big[ \tilde{\ddot{d}}^{(\alpha)}_z(\omega) \big],
\label{eq:PhaseDef}
\end{equation}
the spectral phase of the dipole acceleration of a sample in interferometer arm $\alpha$ and $\Delta\theta_{\text{opt}}$ is the difference in optical path between the two beams.
The argument of the $\cos$ term in Eq.~\eqref{eq:Iint} is a function of the relative phase between the dipole distributions within each sample which contains the fingerprints of the quantum dynamics triggered by the incident fields.
This difference represents the experimentally measured change in phase when some parameter e.g., relative CEP phase, time delay, etc, of the field interacting with one of the sources is varied.
Therefore, we argue, single-photon measurements like the ones presented here can yield information, through photon counting, on the field, and the medium.

To start with BG beams, the stability of the interferometer is thoroughly characterized.
This series of experiments can be split into two measurable quantities: the degree to which the phase delay can be controlled and the resolving power in the phase measurement.

To demonstrate control, one fundamental beam is delayed with respect to the other by using the SLM to add an optical phase which can be tracked in the interference fringes.
Fig. \ref{fig:stability}(a) shows a delay phase scan over \(\pi\) rads of the 17th harmonic. 
The inset of Fig. \ref{fig:stability}(a) shows a scan with steps close to its resolution. 
Here, 7th harmonic pulses are delayed in 0.2~mrad steps over 2~mrads of the fundamental achieving a resolution of $\, 52\,$~zs. 
In other words, we can control two pulses that are spatially separated by 0.3~mm at 72~nm center wavelength (17.3~eV) with a precision that is only found in optical cavities. Overall, this constitutes an improvement of two orders of magnitude in step size compared to earlier experiments \cite{tross_interferometer} and one order of magnitude compared to other state of the art XUV-XUV delay lines as well as phase gratings \cite{jansen2016spatially, camper_high_2019, mandal2021attosecond, koll2022phase}.

The resolving power of these beams is shown in Fig. \ref{fig:stability}(b).
Here the phase stability is reported through three metrics (see Methods section), the standard deviation, $\sigma_s$, the Allan deviation, $\sigma_A$, and the differential deviation, $\sigma_D$.
The standard deviation decreases for ~6 hours and reaches a value of ~18\,zs for the 9th harmonic before long-term drifts begin to dominate and increase the error.
In the Allan deviation we see a small peak at 20 seconds which corresponds to the timescale our beampointing stabilization operates at, as it measures and corrects the position and angle of the laser every 30 seconds allowing some small drift beforehand.
After that, it has a minimum value of ~3.5\,zs for the 9th harmonic before gradually climbing as long-term drifts come into play.
We see $\sigma_D$ drop to ~9\,zs over the course of a day with little sign of stopping, indicating that an even longer measurement would be possible and give more precise results.

Most attosecond experiments have a strong caveat, the strong fundamental co-propagates with the XUV pulses. We again emphasize that this is not the case here. We use tilted BG beams \cite{Rodnova2020, davino2021higher} leading to two strong-field-free XUV beams without the use of filters which would limit the intensity of the XUV beams.
Additionally, one has to mention that with a higher repetition rate source, e.g. 50~kHz which is not uncommon for XUV sources \cite{wang2015bright} or our own XUV source at 200~kHz\cite{watson_high_power_2025}, millions of shots could be reached within seconds. Our laser operates at 1~kHz and reached 3.5~zs after $\sim10^8$ shots.
Ytterbium based sources can even reach repetition rates over 10 MHz \cite{hadrich2015exploring} and could decrease the measurement time to a matter of minutes. Additionally, modern LCOS SLM technology is no longer a limitation as they can handle high powers
\cite{zhu2018investigation, kaakkunen2014fast, carbajo2018power}.

The high stability of the source allows interferometric experiments to run over ling periods of time, opening the door for photon counting experiments in the XUV that make up attosecond pulses. This regime is reached by lowering the generating gas pressure \cite{Shiner2009,Shiner2013} and changing the detector operating regime.

\begin{figure}[ht]
\centering
\includegraphics[width=0.9\textwidth]{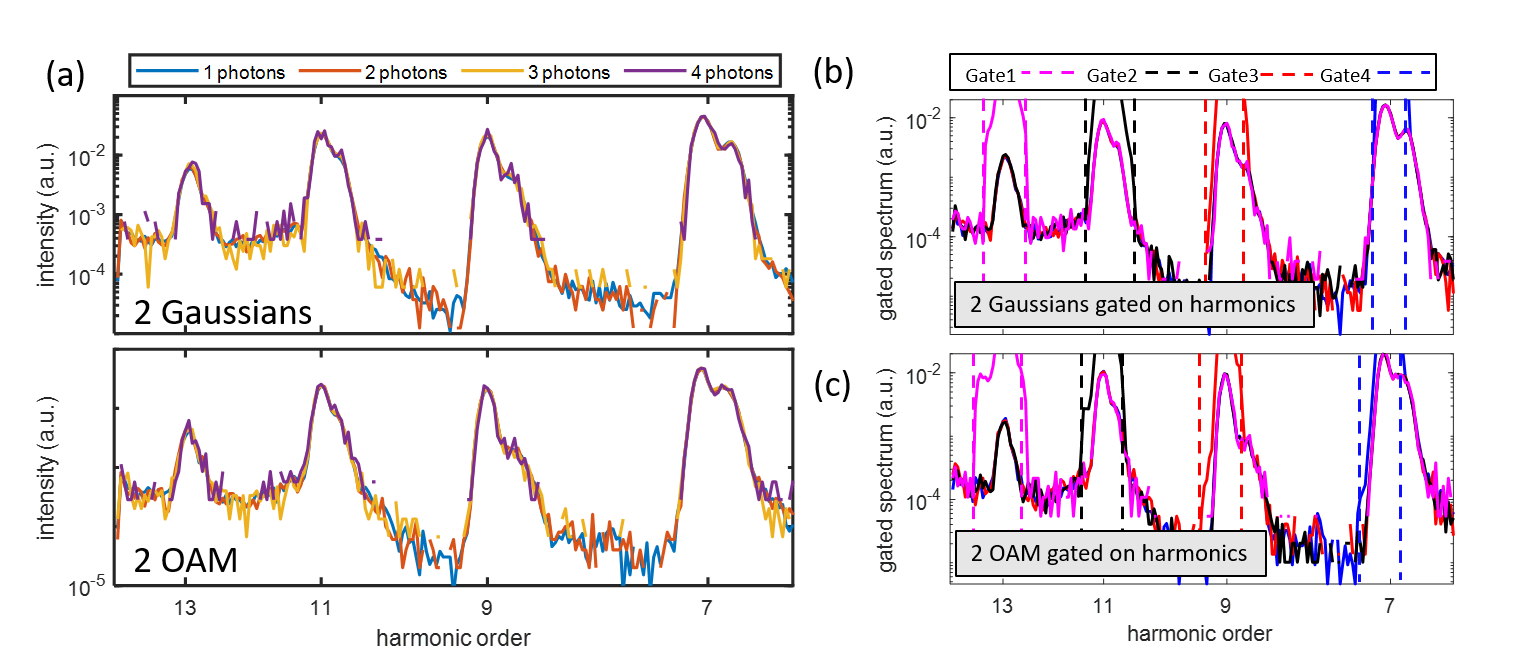}
\caption{
(a) Measured XUV spectra after conditioning for a certain number of photons in the data for 2 beam profiles: 2 Gaussian foci and 2 foci formed by the interference of beams with $l=\pm 1$ OAM. (b) Spectra formed from spectrally conditioning the same beam profiles. The gated harmonic is strongly amplified, yet the rest of the spectrum is identical.
}\label{fig:spectra}
\end{figure}

Initially, correlations between harmonics are inspected by spectrally conditioning the data as described in the methods.
Spectra conditioned on harmonics 7, 9, 11, or 13 are shown in Figure \ref{fig:spectra} (b).
As expected, the signal within the conditioned harmonic is amplified but outside of that no changes to the spectra are found. 
This shows no spectral correlations between the harmonics which supports the recent theoretical work finding that each photon generated through HHG carries the full harmonic spectrum\cite{gorlach2020quantum}.
A more extensive covariance analysis can be found in supplementary section \ref{sect:Correlations}.

Additionally, figure \ref{fig:spectra}(a) shows the harmonic spectra from one or two Gaussian beams conditioned to have 1, 2, 3, or 4 photons. 
It is evident that the spectra do not depend on the number of photons detected indicating that spectral content is independent of the number of photons generated.
To our knowledge, this is the first demonstration of the theoretical proposal on this regard \cite{gorlach2020quantum}. Our experiment indicates that each photon generated through HHG indeed contains information about the entire XUV spectrum.

\begin{figure}[ht] 
\centering
\includegraphics[width=0.7\textwidth]{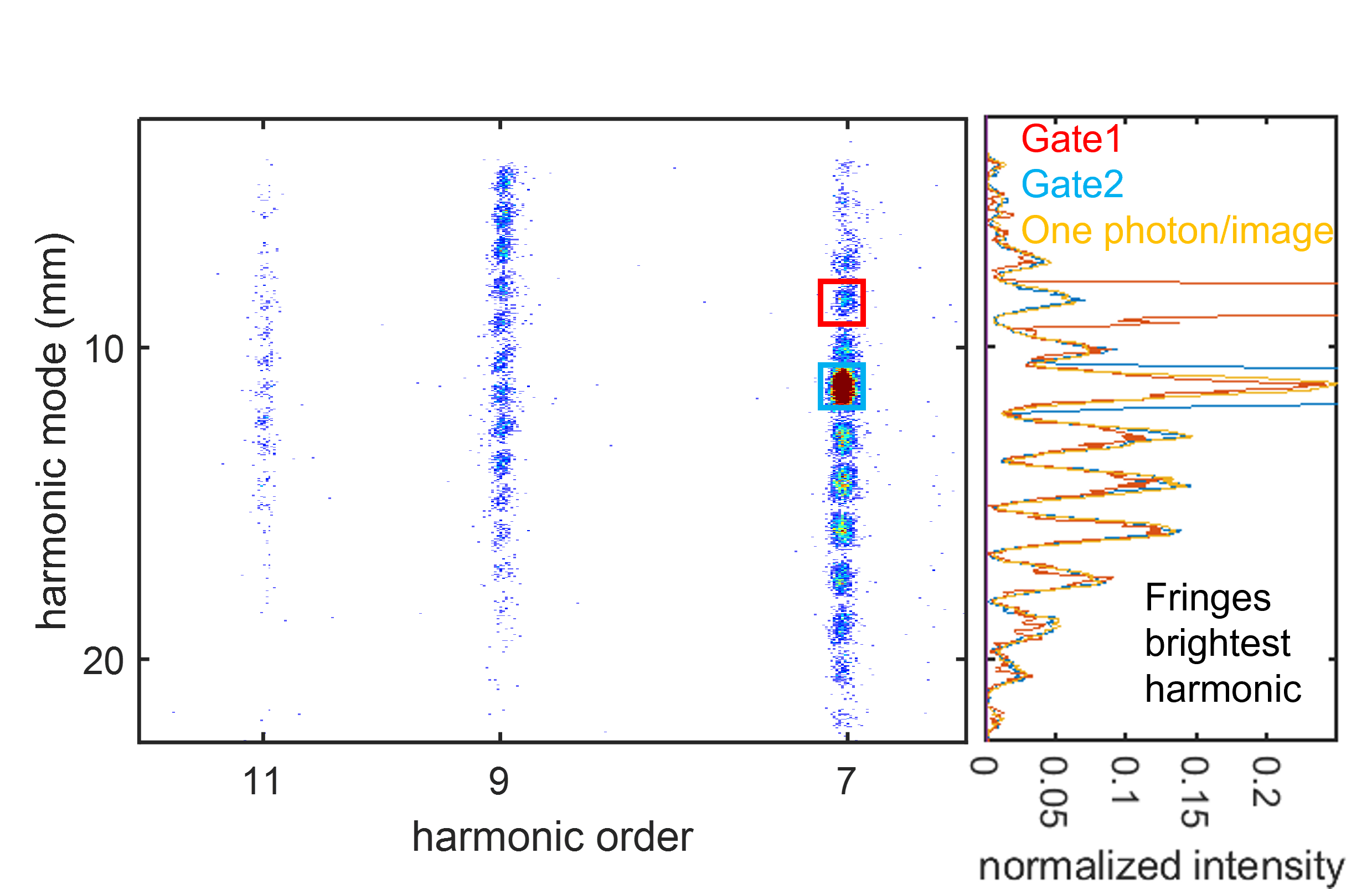}
\caption{Accumulated image imposing two conditions on the detected photons: First, exactly two photons are detected within one image. Second, one photon is detected within a predefined gate. The gate for this image is depicted in blue. Right side shows the projection of the brightest harmonic onto the y-axis for two different gates and for the  single-photon data. The gates used for these fringes are indicated in the image in red and blue.
}
\label{fig:gated_fringe}
\end{figure}

We also need to rule out any spatial correlations between the two beams in order to use pairs of XUV photons for quantum optical experiments. 
These correlations could result from entanglement between photons requiring momentum conservation. 
Photons with different angled momenta are detected at different spatial locations along the harmonics and correlations in their momenta should result in correlations within the fringe patterns. 
For this purpose photon counting data is spatially conditioned as described in the methods. 
Projections of the fringe pattern for the 7th harmonic (113nm) are shown in the right panel of Fig. \ref{fig:gated_fringe} for three cases: 
fringes reconstructed one photon at a time (orange) or with two different spatial conditionings (blue and red).
If the two detected photons were entangled, when collapsing the wavefunction of one, the interference pattern on the detector would get lost, as we should not be able to determine the relative phase of both photons.
Fig. \ref{fig:gated_fringe} clearly shows that the interference fringes do not change when two-photon measurements are done and we collapse the wavefunction one photon at a time.
The overall results show it is now possible to perform precise quantum optical measurements in the XUV with phase precisions that are normally only attainable inside optical cavities.

In our numerical simulations detailed in the supplemental section \ref{sec:numerical_simulations}, we use Argon as a prototype. The multi-electron dynamics in Argon are treated in the context of the Time-Dependent Configuration Singles (TDCIS) which approximates correlated multi-electron dynamics as a linear combination of channel-resolved correlated electron-hole excitations from the neutral Hartree-Fock ground state to initially unoccupied Slater determinants describing single electron-hole excitations~\cite{Greenman2010}.

\section*{Discussion}
\label{discussion}

If the optical phase is held constant, only dipole contributions will be present in the measured phase. A thorough calculation demonstrating this is shown in the supplemental section \ref{sec:numerical_simulations}.
Of special interest are cases where the induced dipole is controlled by a third beam that could be weak (perturbative) or intense(non-perturbative).
This opens the door for interferometric transient absorption spectroscopy methods, where the observable is not restricted to only the absolute value of the dipole but can also measure the phase. Thus, we could access the phase of a quantum wave packet, whether bound or in the continuum. An example is the dynamical study of molecular ro-vibrational and/or electronic wavepackets. In both cases, one or more pulses populate a superposition of excited states, creating wavepackets with an overall phase. Finally, with a precision of 3.5~zs, we could perform experiments to test temporal aspects of QED. For example, tests of the radiation reaction time at near the Compton characteristic timescale. Another possible test at the Compton timescale is the existence of zitterbewegung, the predicted oscillatory motion of the electron as described by the Dirac equation \cite{breitInterpretationDiracsTheory1928,schrodingerFreeMovementRelativistic1930,greinerRelativisticQuantumMechanics1995}. While zitterbewegung has been observed in analog systems \cite{rusinZitterbewegungElectronsGraphene2008,luanZitterbewegungNewDirac2018,lovettObservationZitterbewegungPhotonic2023}, there are currently only compatible experiments for free electrons \cite{gouanereExperimentalObservationCompatible2008,remillieuxHighEnergyChannelling2015}. On the other hand, intense laser pulses and interferometric techniques, like the one presented here, are suitable for precise measurements to rule out the existence of such phenomena \cite{romanZitterbewegungDiracElectron2003}.

Another possible application of our separated beam interferometer is 3D imaging of small samples by measuring changes in the fringe patterns due to differences in optical paths as the sample is moved through the beam.
To show how our interferometric technique could be used, we start with Eq. \ref{eq:Iint} and in particular, the term $\Delta\theta_{\text{opt}}(\mathbf{r}_d,\omega)$.
Contributions to the geometrical phase arise from the difference in optical paths from either source.
So far, we have assumed that XUV pulses travel through the same medium and that the interacting medium has an identical composition.
However, this assumption does not need to hold. If instead, source 1 travels through a medium with an index of refraction $n_1$ and thickness $L^1_S$, while source 2 travels through $n_2$ and $L^2_S$, then our interferometer will measure a phase difference of $\Delta\theta_{\text{opt}}(\mathbf{r}_d,\omega)=-\omega/c\times(n_1 L^1_S-n_2 L^2_S)$.
For 3D imaging of a sample that scans across the XUV beam 2 under vacuum, the phase difference reduces to $\Delta\theta_{\text{opt}}(\mathbf{r}_d,\omega)=(\omega/c) n_2 L(x,y)$, where $L(x,y)$ is the thickness profile of the sample being imaged.
With a phase precision of 0.08~mrad in the 11th harmonic (72~nm), we could measure thickness profiles with a resolution of $1\times 10^{-3}$~nm = 1~pm.
While images taken in this manner only provide a convolution of the optical mode profile and the sample, obtaining relative depth profiles with this resolution could be revolutionary.

Another possibility of using two quantum equivalent XUV photons is exploring true non-local measurements in solids through photoelectron spectroscopy, specifically time-resolved angle-resolved photoelectron spectroscopy (trARPES) \cite{Smallwood2012,Corder2018,Sie2019}.
Theoretical proposals to use two-electron ARPES measurements to uncover correlations have been discussed in the past \cite{Berakdar1998TwoElectrons, Mahmood2022TwoElectrons}.
However, the proposed measurements only involved local electron pairs. Our non-local equivalent XUV photons could probe correlations in a solid across macroscopic distances while preserving the energy and momentum resolution of ARPES. We argue that this could open the door to a new set of experiments to explore and resolve the spatial and temporal extent of quasi-particle correlations.

\section*{Methods}

\paragraph{Experimental}
790 nm, 2.5 mJ, 35 fs pulses from our titanium-sapphire laser were separated by a XY Phase Series 512L Meadowlark SLM into two beams that are focused through a 150 mm lens and 0.5$^{\circ}$ axicon into two foci in an ethylene gas jet.
They interact with a non-linear medium to generate XUV photons that are detected spectrally and spatially in an XUV spectrometer where their phase difference can be measured through the interference fringes. 
The axicon shapes the IR beams into Bessel-Gauss modes which diverge in the far-field leaving isolated XUV pulses without the use of a filter. 
Though not used in these experiments, a pulsed Even-Lavie gas valve is placed a few centimeters after the HHG where a target can be placed into the beams independent of the IR fundamental. 

\paragraph{Resolution of ITAS}
To demonstrate control, one fundamental beam can be delayed relative to the other by using the SLM to add an optical phase \(\Delta \phi_{SLM}\), which can be tracked in the interference fringes for harmonic 2q+1, following \(\Delta \phi_{2q+1} = (2q+1) \times \Delta \phi_{SLM}\).
For a light pulse traveling in vacuum, an optical phase \(\Delta \phi_{2q+1}\) is equivalent to a  time delay, \(\Delta \tau _{(2q+1)} = \frac{\Delta \phi _{2q+1}}{2\pi} \times \frac{T_{f}}{2q+1}\) with \(T_{f}\) being the fundamental's period.
A series of delays should result in a linear measured phase with slope given by the harmonic order as shown in \ref{fig:stability}(a).
The resolution is determined from the root mean square (RMS) deviation from the intended linear slope and can be expressed as a phase or time.

\paragraph{Measures of Uncertainty}
In this paper we use three different measures of uncertainty in an experiment: the standard error, $\sigma_s$, the Allan deviation, $\sigma_{A}$, and the differential error, $\sigma_{D}$. The standard error is commonly used to make confidence intervals for measurements and is used for the error bars in these measurements along with figure \ref{fig:stability}(b). The Allan deviation is specifically the more commonly used overlapping Allan deviation\cite{riley_handbook_2008}. This deviation calculates the root mean squared error of a measurement as a function of how much data is averaged together and provides a measure of how much noise-sources on different timescales affect the data. Then the differential error is defined as the standard error of the difference between subsequent measurements. This measure is closest to how phase changes due to targets will be measured as a background measurement of the optical phase will be needed in tandem with any target phase change. This uncertainty measure removes long term drifts from the measurement as each pair of subsequent images are combined as a background measurement and a target measurement. 

Mathematically, we consider an ordered set of $N$ subsequent measurements of a single data point, $\{x_i\}$ with $i\in \{1,2,...,N\}$ and mean value $\bar{x}$. The standard error is defined as follows: 
$$\sigma_s^2 = \frac{1}{N^2} {\sum_{i=1}^{N} \left( x_{i} - \bar{x}\right)^2}$$

For the Allan deviation we first take the average value of sections of data as $y_i$ which is the average of values from $x_i$ to $x_{i+n}$. Then the Allan deviation is defined by: $$\sigma_A^2(n) = \frac{1}{2 \left( N -2n +1 \right)} {\sum_{i=1}^{N-2n+1} \left( y_{i+n} - y_{i}\right)^2}$$

For the difference error we first take the difference between subsequent measurements, $w_{i} = x_{2i+1} - x_{2i}$, and then check the standard error of that quantity: $$\sigma_D^2 = \frac{4}{N^2} {\sum_{i=1}^{N/2} \left( w_{i} - \bar{w}\right)^2}$$.

\paragraph{Conditioning of Photon Counting Data}
In this work we conditioned the photon counting data in three ways, spatially, spectrally and based on the photon number. The photon number conditioning is simplest as it just involves limiting the dataset to measurements which detected a certain number of photons as in figure \ref{fig:spectra}(a) where spectra are shown from data with 1, 2, 3, or 4 photons.
The spectral conditioning is twofold. First it requires limiting the data to just 2 photon measurements. 
Then a second condition is enforced, that one of the photons measured must have a wavelength within a certain range or "gate". This is used in figure \ref{fig:spectra}(b) where spectra are shown from 2-photon data with one of the photons being in harmonic 7, 9, 11, or 13.
This second requirement results in an amplification of the signal within the gate because at least half the data must be inside that region. However, since this only effects one harmonic in each conditioning there are still 3 spectra to compare to each other at each harmonic.  
The spatial conditioning used here is most limiting of all since it first involves a 2-photon condition and then limiting the data to measurements where at least one photon was detected in a region or "gate" of our detector. Figure \ref{fig:gated_fringe} uses this method with the gates outlined in red or blue squares.
For these tests data is gated on a single fringe of harmonic 7, though the results are unchanged if gated differently. 
As with the spectral conditioning, this amplifies the data within the gate but regions outside of it can be compared.

\section*{Data Availability}
Data available on request from the authors.


\section*{Acknowledgements}
This work was done under US US Department of Energy, Office of Science, Chemical Sciences, Geosciences, \& Biosciences Division grant DE-SC0024508. ATL and REG were supported by the U.S. Department of Energy (DOE), Office of Science, Basic Energy Sciences (BES) under Award Number DE-SC0023192.
M.~F.~C.~acknowledges support by the National Key Research and Development Program of China (Grant No.~2023YFA1407100), Guangdong Province Science and Technology Major Project (Future functional materials under extreme conditions - 2021B0301030005), the Guangdong Natural Science Foundation (General Program project No. 2023A1515010871), and the National Natural Science Foundation of China (Grant No. 12574092). 

\section*{Author contributions statement}

GRH and TS performed the experiments, analyzed the data, and interpreted the results. CT-H conceived the experiments and helped interpret the results. George Gibson helped with the interpretation of the results. ATL and REG did the theoretical calcuations for the two source interference. CG, BKD, and MFC did the calculations for the theoretical calculations two Bessel

\section*{Competing Interests}
The authors declare no competing interests.

\newpage








{\raggedright\sffamily\bfseries\fontsize{20}{25}\selectfont  Single photon zeptosecond interferometry: supplemental document \par}%
\vskip18pt%

\section{Phase Mask Creation for Gaussian Beams} \label{subsect:Supple_gaussian_beams}

\begin{figure}[ht]%
\centering
\includegraphics[width=320pt]{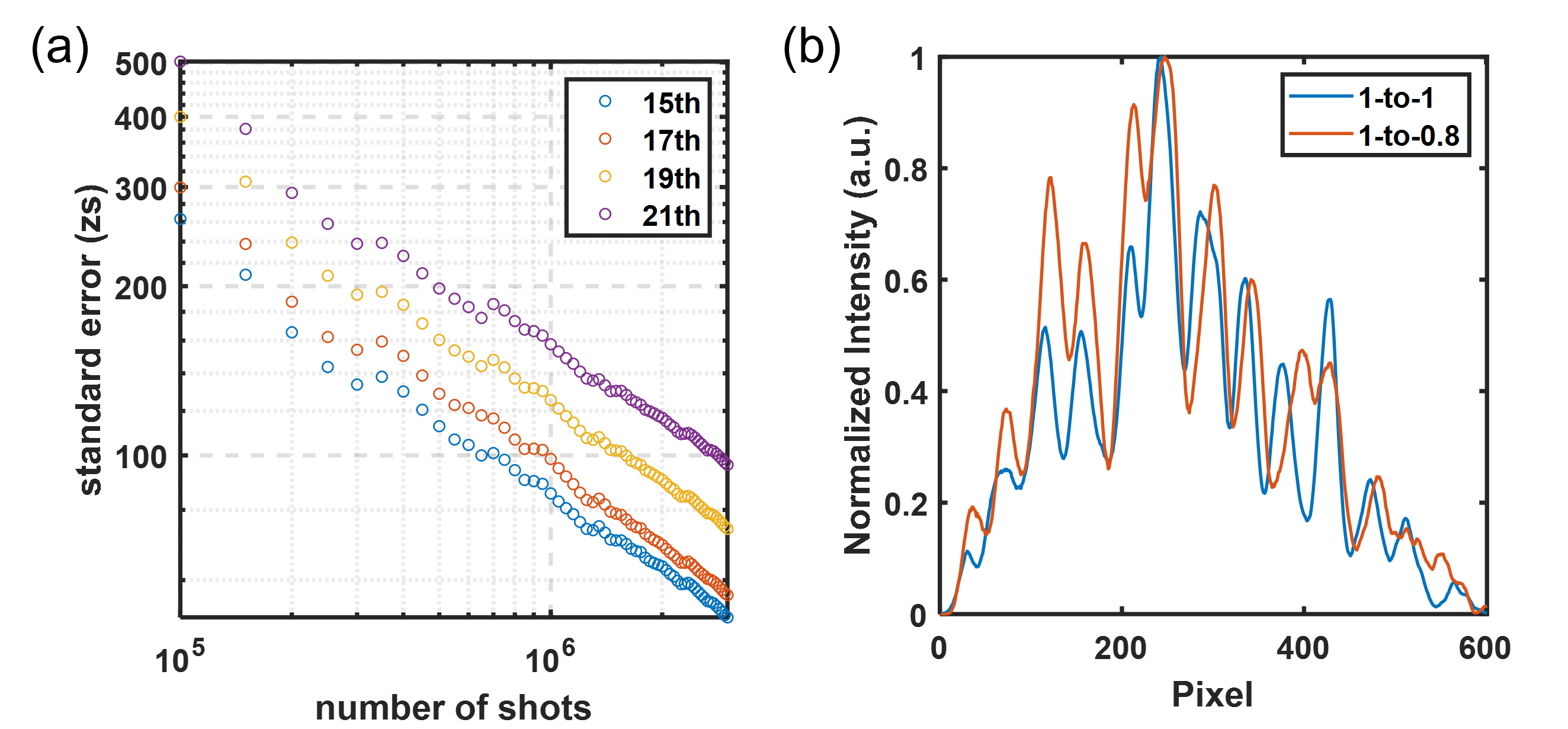}
\caption{(a) Standard error in zs of two balanced Gaussian beams as a function of the number of shots. Data is shown for the 15th - 21st harmonic. (b) Unbalanced fringes for the 7th with the fundamental beam split 1-to-1 and 1-to-0.8
} \label{fig:gaussian_std_error_and unbalanced}
\end{figure}

The multiple beam phase masks for the SLM are generated by intertwining two opposing linear slopes.
This is done by randomly assigning each pixel to a particular mask.
The randomization prevents structures from forming increasing the diffraction efficiency of the device.
It also allows us to arbitrarily distribute the light between the beams by biasing the random number assignment, for example giving one beam 50$~\%$ of the pixels and the other 40$~\%$ would give a 20$~\%$ difference between the intensities at the focus, which corresponds to an unbalanced scheme. In such a case, the extra ten percent is put in a third beam that focuses far away from the others and doesn't affect the experiment.
In general, this method is not limited to two beams but can be scaled depending on the pixel resolution of the SLM and the beam size impinging on it. 
The stability of a balanced scan (15th - 21st harmonic) as well as a sample line cut of one harmonic (7th) for the unbalanced case is given in figure \ref{fig:gaussian_std_error_and unbalanced}(a) and (b), respectively.
Even within only $3\times10^6$ shots these harmonics reach a precision of under 100-zs.

\section{Orbital Angular Momentum beam generation}
\label{sect:supple_oam}

To check for entanglement beams with superimposed $\pm 1$ units of orbital angular momentum (OAM) are created in the following scheme, as presented before by Tross et al. \cite{tross2019high}.
In the same manner as for the Gaussian beams pixels were chosen at random and one of two different phase masks is applied. The first phase mask continuously ramps up the phase circularly from 0 to 2$\pi$, whereas the other starts at 2$\pi$ and decreased circularly to 0.
It is important that the phase jump between 0 and 2$\pi$ happens at the same spot on the SLM. The resulting co-linear beam is a superposition of OAM order $l=\pm 1$  and when focused by a lens develops two distinct foci, which can be used for HHG.

\section{Bessel-like beam generation}
\label{sect:supple_bessel_gen}

\begin{figure}[ht] 
\centering
\includegraphics[width=250pt]{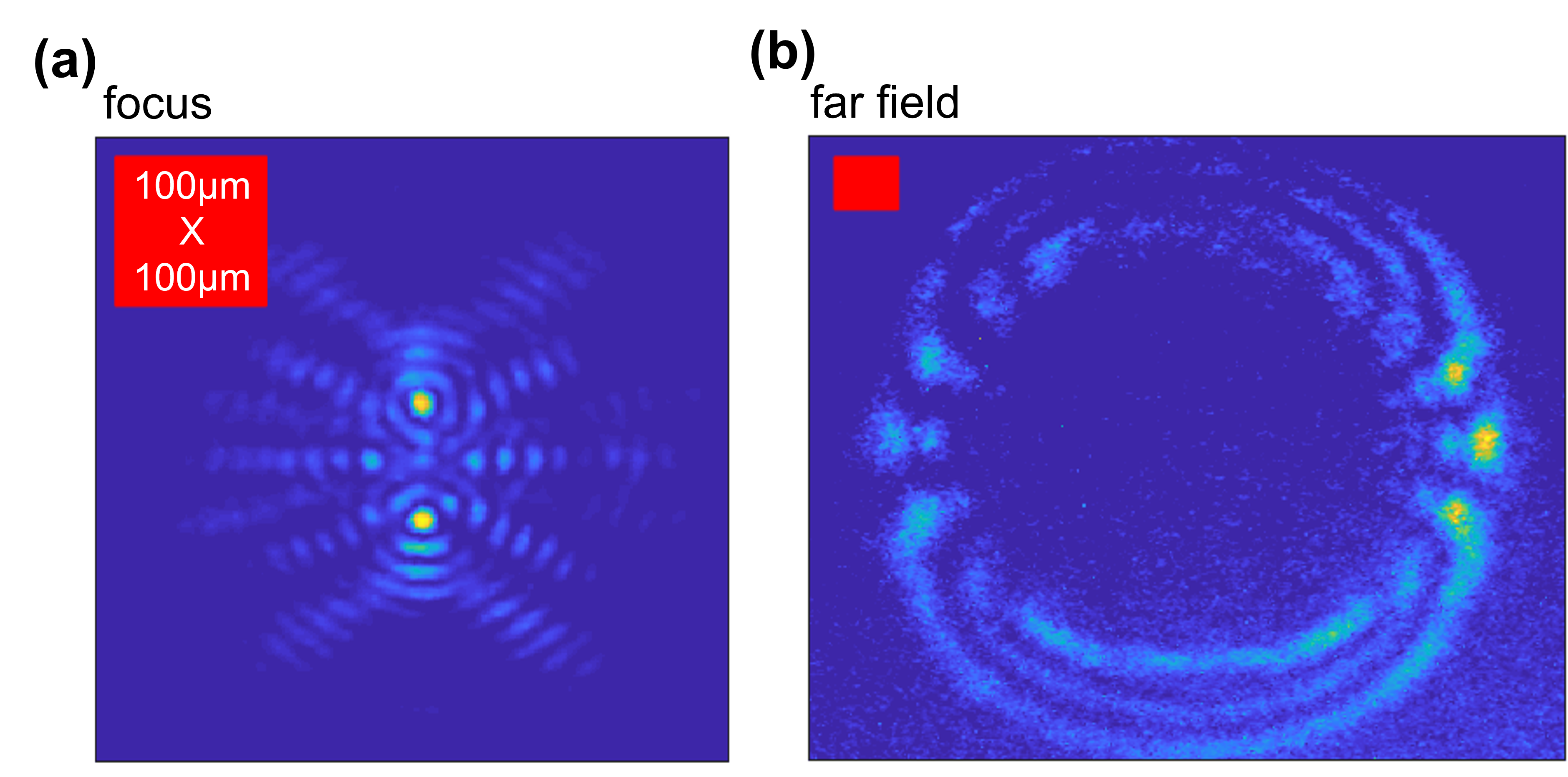}
\caption{
(a) The focus of two Bessel-like beams with two different interwoven wavefronts. (b) Far field of the beams depicted in (a).
}\label{fig:bessel}
\end{figure}

To switch from Gaussian beams to Bessel-like beams only one optical element needs to be added to the optical system: an axicon after the lens. Such beams have been used in the past for HHG in a thin gas jet \cite{davino2021higher}.
Figure \ref{fig:bessel} (a) shows a typical focus achieved when using these beams in conjunction with two different wavefronts as in the Gaussian case.
Figure \ref{fig:bessel} (b) shows the far field with zero intensity on axis, well suited for IR-free experiments.

\section{Phase evaluation of the experimental data}
\label{sect:phase_evaluation}

\begin{figure}[ht] 
\centering
\includegraphics[width=250pt]{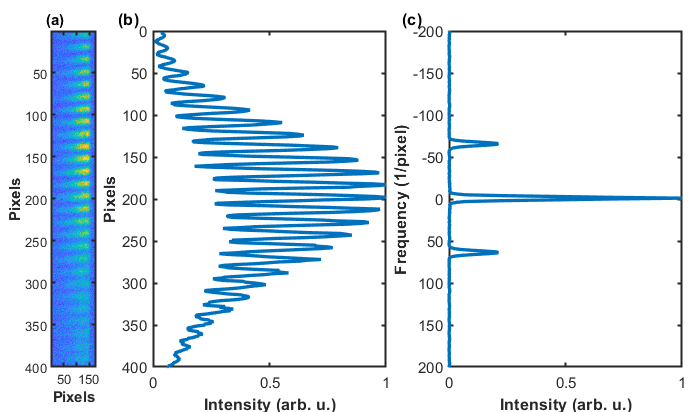}
\caption{
(a) A zoomed in harmonic showing fringes.
(b) An integrated lineout of the fringe in (a) with a hamming filter applied. For the analysis this line is also zero-padded but the padding is cropped out for clarity.
(c) The Fourier Transform of the lineout shown in (b). This figure is again cropped for clarity. The only peaks present are those corresponding to the DC component and the frequency of the fringes.
}\label{fig:FFT}
\end{figure}

To extract reliable phase information from the collected images several steps are taken.
First, each harmonic gets projected on the spatial axis such that the fringes are obtained for each (a sample for one harmonic is given in figure \ref{fig:FFT}).
Second, the projected harmonic has a constant background subtracted.
Third, a Hamming filter is applied to suppress windowing effects and the fringes are zero padded with $2^{10}$ zeros on each side.
As a forth step the fast Fourier-transform (FFT) of the data is calculated, see figure \ref{fig:FFT} (c) for the magnitude of the FFT.
The fringes are clearly visible as a distinct frequency.
The phase of this frequency gives the phase difference between the two beams up to a constant offset. The offset depends on where in a cycle the fringes are cut off by the detector edge.

\section{Full Correlations Between Harmonics}
\label{sect:Correlations}

\begin{figure}[ht] 
\centering
\makebox[\textwidth][c]{\includegraphics[width=1.2\textwidth]{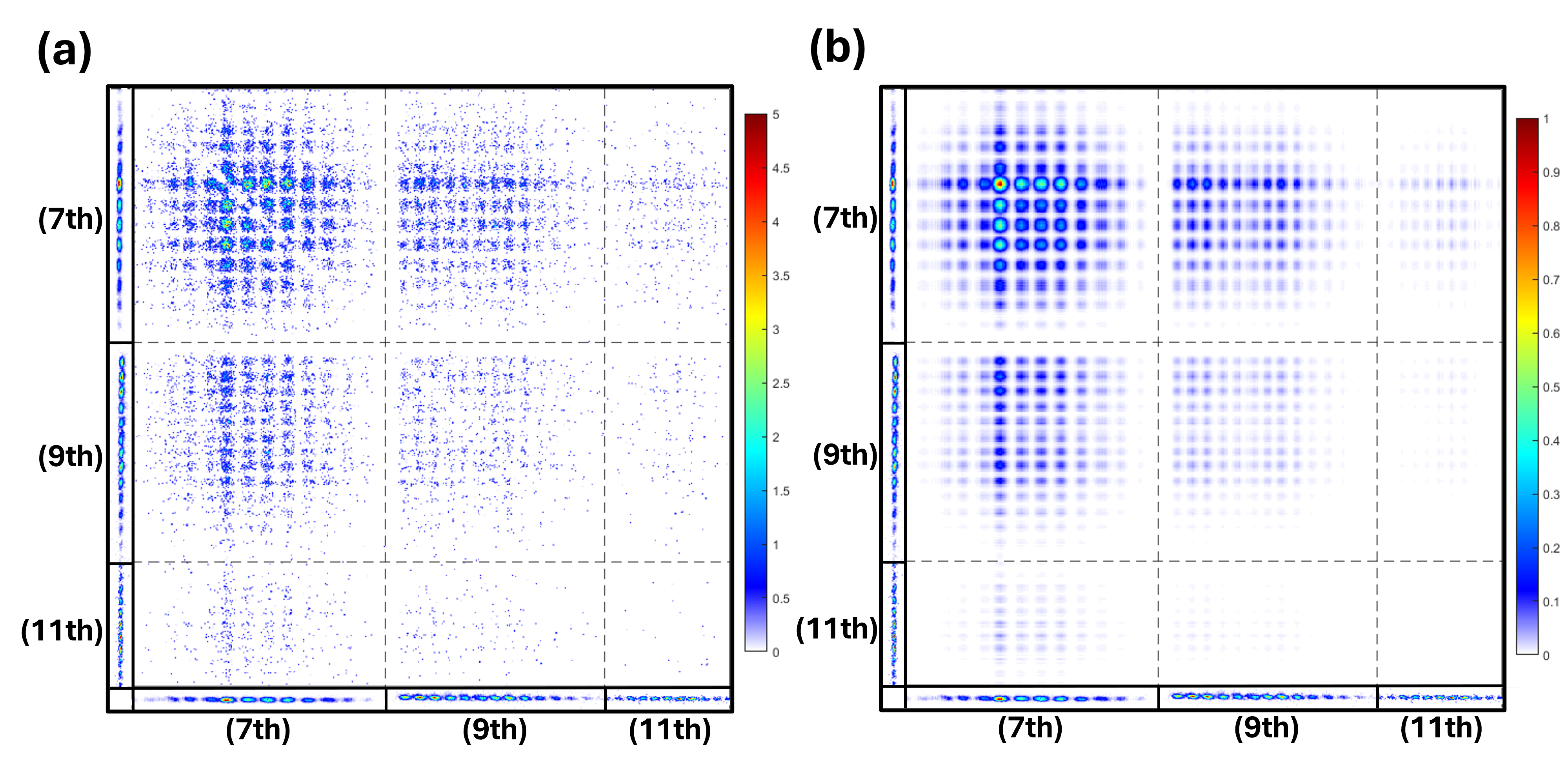}}%
\caption{
(a) Here we show complete coincidence structure between the harmonics shown in figure \textcolor{red}{\ref{fig:gated_fringe}}. Each count in the image represents two hits. One in the harmonic matching it's y-axis position and one in the harmonic matching it's x-axis position. This measurement is sensitive to both spectral and spatial correlations simultaneously. 
(b) This shows an expected coincidence structure reconstructed from single hit data. It shows the structure expected in (a) simply due to the relative amplitude of the harmonics and the fringes in each harmonic. 
}\label{fig:correlation_full}
\end{figure}

In section the main text the spectral and spatial correlations between harmonics were discussed. However, the full correlations between spectral and spatial components were not discussed. In figure \ref{fig:correlation_full}(a), the full coincidence structure of the harmonics from figure \textcolor{red}{\ref{fig:gated_fringe}} is shown. For this measurement only two photon events are considered and the position of one hit in its harmonic is used as the x-axis position and the position of the other hit gives the y-axis position; the harmonics are displayed along the x and y axes for reference. By this definition there is a symmetry across the diagonal and a correlation between harmonics would show as an amplification (or reduction) of one of the off-diagonal dashed boxes. Additionally in this figure correlations between the top and bottom beams could also be seen, as amplifications in the regions corresponding to the top half of one harmonic and the bottom of another.

Figure \ref{fig:correlation_full}(b) shows a reconstruction of the expected coincidence structure of the harmonics based on single-hit data in the fringes. This shows the basic structure expected in (a) simply due to the structure of the fringes. Any amplification or suppression in (a) can be observed relative to these values. 

No clear correlations are observed in the harmonics shown or in any others discussed in this paper. The only significant effect seen in (a) is a diagonal line of missing points corresponding to two hits on the same place on the detector. These double hits appear as one hit in the data analysis and so are omitted from this plot.

\section{Theory of Two-beam interferometry}\label{sec:two_beam_interferometry}
\begin{figure}[ht] 
\centering
\includegraphics[width=0.98\linewidth]{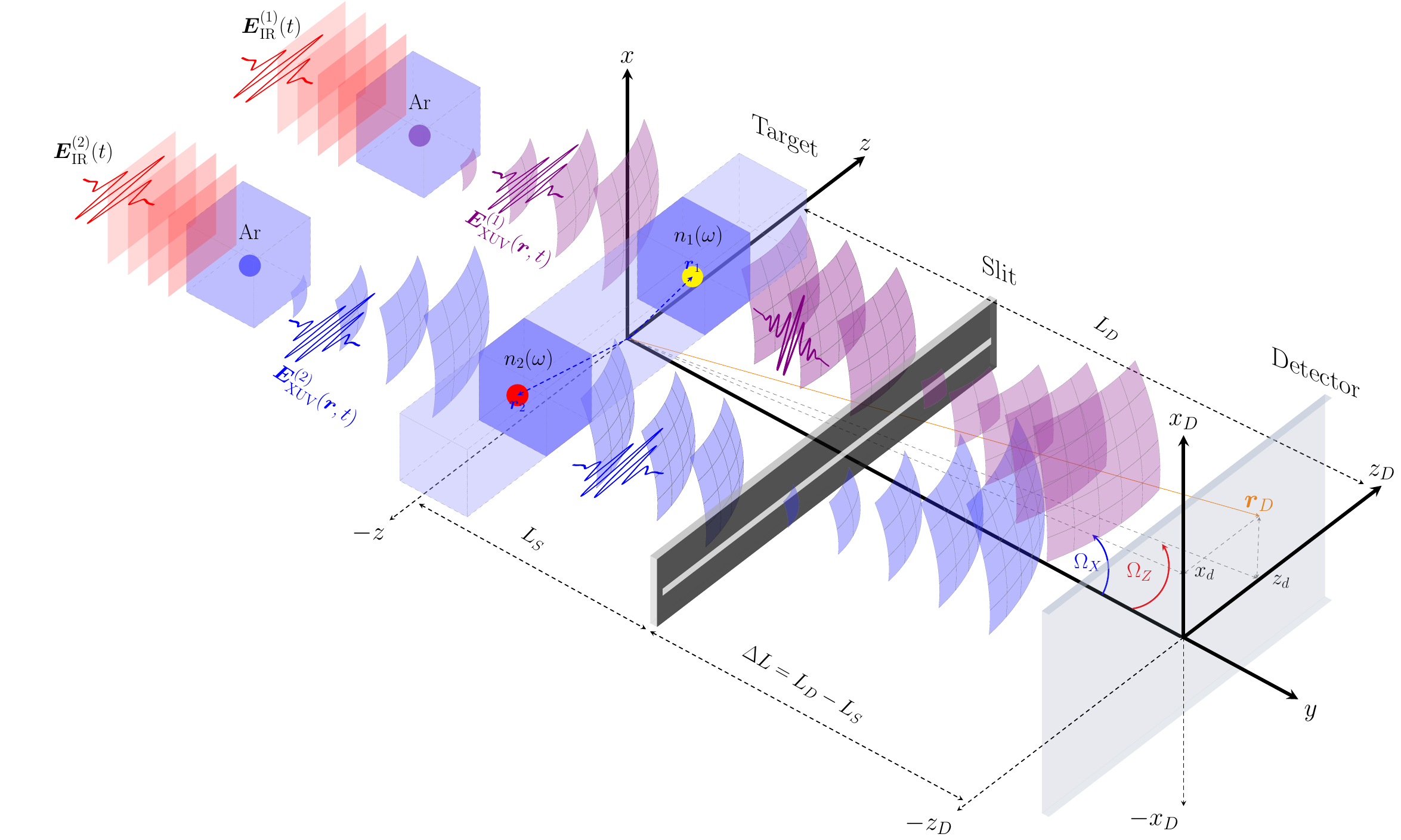}
\caption{\textbf{Idealized interferometric set up}: Two XUV sources $E^{(1)}_{\mathrm{XUV}}$ 
and  $E^{(2)}_{\mathrm{XUV}}$ generated by two phase-locked IR fields through HHG interact with
two spatially separated samples, $S_{1}$ and $S_{2}$, characterized by the indices of refraction $n_1(\omega)$ and $n_2(\omega)$. After interacting with the sources,
and passing trough a diffraction system (Slit) the resulting output fields interfere on the Screen/Detector plane. The output fields resulting from each arm are obtained by solving the Maxwell's equations
at the single-atom level, and are functions of the atomic dipole probed by each XUV field. Interference between both fields 
encodes information about the IR relative CEP phase and exhibits oscillations as a function of the spectral phase difference of each atomic dipole probed by the incident XUV fields on each sample.
} \label{fig:scheme}
\end{figure}

An illustrative sketch of the two-pathway interferometric setup is shown in Fig.~\ref{fig:scheme} setup. As the experimental observable, we consider the time-averaged flux of the Poynting vector~\cite{born_wolf_2019,saleh_teich,siegman,Shore1989} arriving at the detector plane, i.e., 
\begin{eqnarray}
\label{eq:averaging}
\langle \boldsymbol{S}_D(\boldsymbol{r}_{d},t)\rangle\equiv(1/T)\int \boldsymbol{S}_D(\boldsymbol{r}_d,t)\, dt\,,
\end{eqnarray}
where $\boldsymbol{S}_D(\boldsymbol{r}_d,t)\!=\!(\epsilon_0/\mu_0)\, \big|\sum_{\alpha} \boldsymbol{E}^{\prime (\alpha)}(\boldsymbol{r}_d\!-\!\boldsymbol{r}_\alpha,t)\big|^2 \, \text{\textbf{e}}_r$, 
as defined in appendix Eq.~\eqref{eq:Poynting1},
is the instantaneous intensity (directional energy flux) of the EM radiation at a point $\boldsymbol{r}_d$ in the detector plane;
 $\boldsymbol{r}_\alpha=(x_\alpha,y_\alpha,z_\alpha)$ the origin of coordinates associated to the sample $\alpha\!=\!1,2$ with respect to the coordinate system $x,y,z$ in Fig.~\ref{fig:scheme}, 
and $\boldsymbol{E}^{\prime (\alpha)}(\boldsymbol{r}_d\!-\!\boldsymbol{r}_\alpha,t)$ the electric vector field arriving to the detector from the sample $\alpha$ after being diffracted by a system of
slits of aperture function $S(x_s,z_s)$ located at a fixed $y_s=L_S$ with respect to the coordinate system $(x,y,z)$ in Fig.~\ref{fig:scheme}. Expressing 
$\boldsymbol{E}^{\prime (\alpha)}_\vartheta(\boldsymbol{r}_d\!-\!\boldsymbol{r}_\alpha,t)$
in terms of its frequency components, we obtain, after time integration in~\eqref{eq:averaging}, the frequency distribution, 
denoted $\langle \boldsymbol{S}_D\rangle(\boldsymbol{r}_d, \omega)$ and defined by 
$\langle \boldsymbol{S}_D\rangle(\boldsymbol{r}_d, \omega) \equiv d\langle \boldsymbol{S}(\boldsymbol{r}_d,t)\rangle\,/\,d\omega $, which reads,
\begin{eqnarray}
\label{eq:freq_distrib}
\langle \boldsymbol{S}_D\rangle(\boldsymbol{r}_d, \omega)= \big|\sum_{\alpha} \boldsymbol{E}^{\prime (\alpha)}(\boldsymbol{r}_d-\boldsymbol{r}_\alpha ,\omega)\big|^2 \, \text{\textbf{e}}_r\,,
\end{eqnarray}
with $\hat{\boldsymbol{E}}^{\prime (\alpha)}(\boldsymbol{r},\omega)\!\equiv \mathcal{F}\big[\boldsymbol{E}^{\prime (\alpha)}(\boldsymbol{r},t)\big](\omega)\!=\!\int^{\infty}_{\infty} e^{+i\omega t} \boldsymbol{E}^{\prime (\alpha)}(\boldsymbol{r},t)\, dt$.

We consider an slit with aperture function $S(x_s, z_s) = \text{rect}\big(x_d/\Delta_x\big) \times \text{rect}\big({z_d/\Delta_z}\big)$ 
with the aperture widths satisfying $\Delta_x\!\ll\!\Delta_z$, where  $\text{rect}(x_d/\Delta_x)$ denotes a rectangular function 
in $x_d$ and correspondingly for $z_d$. In the reference frame associated to each oscillating charge distribution (samples $\alpha=1,2$), 
a point $\boldsymbol{r}_{s,\alpha}=\boldsymbol{r}_s-\boldsymbol{r}_\alpha$
on the slit aperture parallel to the $y$ (incident) direction is, with the coordinate convention of~\eqref{eq:Efield_final}, 
then determined by $\phi_\alpha\approx\pi/2$ and $\theta_\alpha\approx\pi/2$. Consequently, according 
to~\eqref{eq:Efield_final2}, $E_\vartheta(\boldsymbol{r}_{s,\alpha},t)\approx\mu_0 \ddot{d}_z(t)/4\pi r_{s,\alpha}$. This is, only the $z$-component of the dipole acceleration
contributes to the RHS of~\eqref{eq:Efield_final2}. As for~\eqref{eq:Efield_final3}, the oscillations in the $x$ direction of the dipole acceleration can
be neglected as the leading contribution is given by $\ddot{d}_z$ for an atomic system interacting with a field linearly polarized along the $z$ direction. If this condition 
is not fulfilled, then both, ~\eqref{eq:Efield_final2} and \eqref{eq:Efield_final3} must be used in~\eqref{eq:freq_distrib}. 
Within this configuration and making use of   
$\mathcal{F}\{d(t-r/c)\}(\omega)\!=\!e^{i\omega r/c}\mathcal{F}\{d(t)\}(\omega)$, the dipole acceleration
$\hat{\ddot{d}}^{(\alpha)}_z(\omega)$ and field $\hat{E}^{\prime (\alpha)}_z(\boldsymbol{r}_d,\omega)$ spectra are related, in the Fraunhoffer approximation, according to 
\begin{eqnarray}
  \label{eq:Ew_final}
  E^{\prime(\alpha)}_{z}(\boldsymbol{r}_d,\omega)= -i\frac{\mu_0\omega}{4\pi r_{ds}}\,\,  
  \tilde{\ddot{d}}^{(\alpha)}_z(\omega)\,\, \hat{S}(k^\alpha_x,k^\alpha_z)
  e^{i\frac{\omega}{c}L_{D}}\,\,\, e^{i\frac{\omega}{c}\frac{x^2_d+z^2_d}{2\Delta L_D}}\,\,\,
  e^{i\frac{\omega}{c} n_\alpha L_{S}}\,\,\, e^{i\frac{\omega}{c}\frac{x^2_\alpha+z^2_\alpha}{2L_D}}\,,\quad
\end{eqnarray}
with $\boldsymbol{r}_{\alpha}\!=\!(x_\alpha, y_\alpha, z_\alpha)$ 
the origin of coordinates of the sample $S_\alpha$  
with respect to the coordinate system $(x,y,z)$ in Fig.~\ref{fig:scheme},
$n_\alpha\equiv n_\alpha(\omega)$
the corresponding index of refraction;  
$L_S$ the fixed position of the slit aperture in the $\hat{e}_y$ direction,
$L_D$ the detection plane position in the $\hat{e}_y$ direction; $\boldsymbol{r}_d\!=\!(x_d,y_d,z_d)\!=\!(x_d,L_D,z_d)$ a point in the detector plane
with fixed position at $y_d=L_D$
and finally, $\Delta L_D\!=\!L_D\!-\!L_S$ as indicated in Fig.~\ref{fig:scheme}. Finally, $\hat{S}(k^\alpha_x,k^\alpha_z)$ is the Fourier transform of the
aperture function, given by $\hat{S}(k^a_x, k^a_z) \!=\! \int \int S(x_s, z_s)\, e^{-i k^{\alpha}_{x}(\omega) x_s}
e^{-i k^{\alpha}_z(\omega) z_s} dx_s dz_s$, evaluated at the wave numbers 
$k^{\alpha}_x(\omega)\!=\!(\omega/c)\big(z_d/{\Delta L_{D}}\!+\!n_\alpha(\omega) z_\alpha/L_S  \big)$, 
and 
$k^{\alpha}_z(\omega)\!=\!(\omega/c)\big(z_d/\Delta L_{D}\!+\!n_\alpha(\omega) x_\alpha/L_S  \big)$.

With the exception of $\tilde{\ddot{d}}^{(\alpha)}_z(\omega)$ in \eqref{eq:Ew_final},
all terms contain details of the diffraction and relative path differences (setup geometry and relative position of the samples). 
These relative phases, however, remain unchanged for fixed $\boldsymbol{r}_d$. Consequently, any change in spectral phase in the distribution~\eqref{eq:freq_distrib} can thus 
be attributed to changes in the relative spectral phase bewteen the acceleration dipoles  $\tilde{\ddot{d}}^{(\alpha)}_z(\omega)$  
in \eqref{eq:Ew_final} showing up in \eqref{eq:freq_distrib}, which expands according to 
\begin{subequations}
\begin{eqnarray}
\label{eq:pointyng_freq_distrb}
	\langle \boldsymbol{S}_D\rangle(\boldsymbol{r}_d, \omega)&\!=&\!\boldsymbol{\mathcal{I}}^{(1)}_{\text{DC}}(\boldsymbol{r}_d,\omega) + \boldsymbol{\mathcal{I}}^{(2)}_{\text{DC}}(\boldsymbol{r}_d,\omega) 
	+\boldsymbol{\mathcal{I}}^{(1,2)}_{\text{int}}(\boldsymbol{r}_d,\omega)\,,\quad\nonumber\\
\end{eqnarray}\\[-0.6cm]
with  the DC components,
\begin{eqnarray}
	\boldsymbol{\mathcal{I}}^{(\alpha)}_{\text{DC}}(\boldsymbol{r}_d,\omega) &=& \boldsymbol{\mathcal{I}}_o(\omega)\,\, \big|\tilde{\ddot{d}}^{(\alpha)}_z(\omega)\big|^2\,\, \big|\hat{S} (k^{(\alpha)}_x, k^{(\alpha)}_z)\big|^2\,,\quad\quad
\end{eqnarray}
	for $\alpha=1,2$. $\boldsymbol{\mathcal{I}}_o(\omega)=(\mu_o/4\pi)(\omega/c)^2\, \textbf{\text{e}}_r$ and where
$\boldsymbol{\mathcal{I}}_{\text{int}}(\boldsymbol{r}_d,\omega)$ is an oscillating function resulting from the interference of the output fields of arms ($1$) and ($2$) after the slit,
\begin{eqnarray}
\label{eq:eq_interf}
	\boldsymbol{\mathcal{I}}_{\text{int}}(\boldsymbol{r}_d,\omega) = \boldsymbol{\mathcal{I}}_o(\omega)\!\!\prod_{\alpha=1,2} \big|\tilde{\ddot{d}}^{(\alpha)}_z(\omega)\big|\,\, \big|\hat{S} (k^{(\alpha)}_x, k^{(\alpha)}_z)\big|\,
\cos\big[\phi^{(1)}_{dip}(\omega) -\phi^{(2)}_{dip}(\omega)  + \Delta\theta_{\text{opt}}(\boldsymbol{r}_d,\omega)\big]\,,
\end{eqnarray}
\end{subequations}
with $\phi^{(\alpha)}_{dip}(\omega)\equiv\arg\big[ \tilde{\ddot{d}}^{(\alpha)}_z(\omega) \big] $ the spectral phase of the dipole acceleration, and
$\Delta\theta_{\text{opt}}=\Delta\theta_M(\omega)+\Delta\theta_P(\omega)+\Delta\theta_S(\boldsymbol{r}_d,\omega)$  an optical phase, where
\begin{subequations}
\begin{eqnarray}
	\Delta\theta_{M}(\omega) &=&  \omega \Delta n(\omega)L_S/c\,,
\end{eqnarray}
is a phase resulting from the refractive index difference $\Delta n(\omega) = n_1(\omega)-n_2(\omega)$ of the samples $S_\alpha$. Next,
\begin{eqnarray}
\Delta\theta_P(\omega) =\dfrac{\omega/c}{2\Delta L_D}\left[ (  n_2(\omega)\big(x^2_2 - z^2_2\big) -   n_1(\omega) \big(x^2_1 - z^2_1\big)  \right]\,,
\end{eqnarray} 
is the optical phase arising from the path difference due to the relative positions $\boldsymbol{r}_\alpha$, and finally, 
\begin{eqnarray}
\label{eq:eq_spectral_phase_abs}
\Delta\theta_S(\boldsymbol{r}_d,\omega) = \theta^{(1)}_S(\boldsymbol{r}_d,\omega)  -\theta^{(2)}_S(\boldsymbol{r}_d,\omega)\,,
\end{eqnarray}
\end{subequations}
with $\theta^{(\alpha)}_S(\boldsymbol{r}_b,\omega)=\arg \big[ \hat{S}(k^{(\alpha)}_x, k^{(\alpha)}_z)\big]$ 
the optical phase at the detection point $\boldsymbol{r}_d$ due to diffraction by the slit.

\begin{figure}[ht]%
\centering
\hspace{-0.5cm}\includegraphics[width=0.50\linewidth]{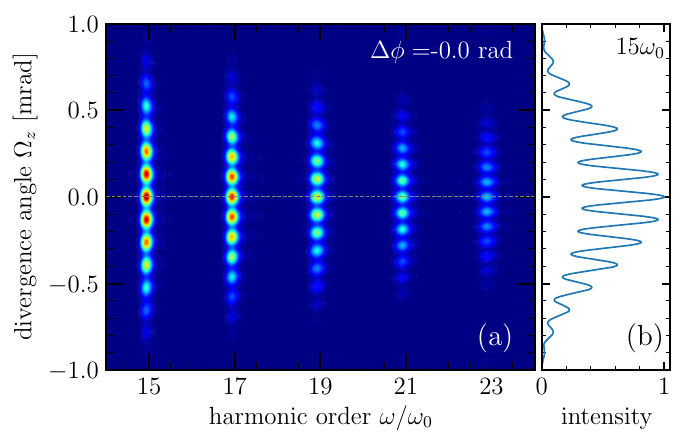}\includegraphics[width=0.50\linewidth]{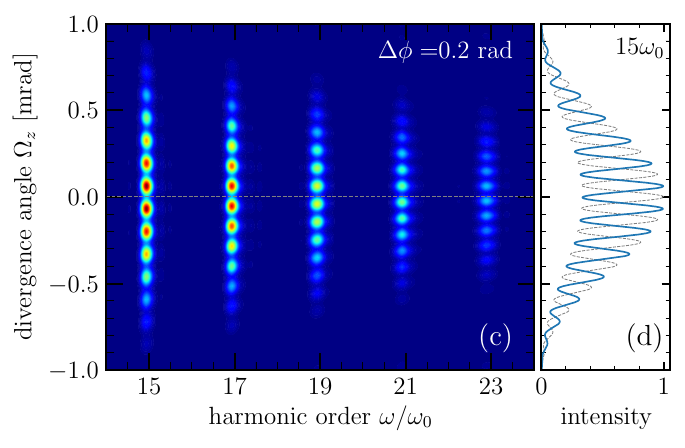}
\caption{
\textbf{Numerical Simulations}: (a) Frequency distribution of the volume-integrated HHG signal as a function of the divergence angle $\Omega_z$ defined in Fig.~\ref{fig:scheme} for $\Delta\phi=0$ rad. (b) Normalized intensity profile of harmonic $15$ for $\Delta\phi = 0$. (c) Same as (a) but for $\Delta\phi=0.2$ rad. (d) Intensity profiles for $\Delta\phi=0.2$ rad (solid-blue line) and $\Delta\phi=0$ rad (dashed-gray line) shown as a reference. Intensity profiles in (b) and (d) are normalized with respect to the peak intensity in (b). As a function of $\Delta\phi$, the intensity profiles of the harmonics are shifted with respect to the reference signal, i.e., compare intensity profiles shown in blue and gray colors in panel (d) . These shifts are used to extract the relative spectral phase of the atomic dipoles shown in Fig.~\ref{fig:dp_vs_cep_supple}(b). Geometry parameters are $x_\alpha=0$, $y_\alpha=0$ and $z_\alpha=\pm 50\,\mu\text{m}$ for 
the position of the samples $\alpha=1,2$; and $\Delta x = \Delta y  = \Delta z = 50\,\mu\text{m}$ for their dimensions in the $x,y,z$ direction.
} \label{fig:numerical_fringes1}
\end{figure}

\section{Numerical Simulations} \label{sec:numerical_simulations}
\subsection{Quantum Dynamics and Multi-electron wave function}

The spectral distribution $\hat{\ddot{d}}^{(\alpha)}_z(\omega)$
in~\eqref{eq:Ew_final} is obtained from the Fourier Transform of the quantum mechanical dipole acceleration
$\ddot{d}^{(\alpha)}_z(t)= \boldsymbol{e}_z\cdot\langle\Psi^{(\alpha)}(t)|-\nabla\hat{\mathrm{U}}(\boldsymbol{r})|\Psi^{(\alpha)}(t)\rangle$
where $|\Psi^{(\alpha)}(t)\rangle$ designates the wave function propagated under the influence of the field-free Hamiltonian $\hat{H}_0$ and the external IR field in arm $\alpha$
in the dipole approximation, as indicated above and $\mathrm{U}(r)$ is the combined atomic and electric field potential energy functions. We simulate the experimental HHG measurements using Argon as a prototype. The multi-electron dynamics 
in Argon is treated in terms of the Time-Dependent Configuration Interactions 
Singles (TDCIS) with active $3s$ and $3p$ orbitals. The TDCIS formalism describes 
single excitations of correlated electron-hole pairs from the Hartree-Fock ground state to initially unoccupied Hartree-Fock orbitals~\cite{Greenman2010}. For 
each arm $\alpha=1,2$, the IR pulse in~\eqref{eq:E_ext} is modeled by a 800 nm (central frequency $\omega_0=1.55\,$ eV), transform-limited pulse of full 
width at half maximum in intensity of 30 fs with Gaussian envelope and fixed carrier-envelope phase (CEP). The CEP in arm 1 is fixed to zero while the CEP if the IR field
acting on the sample $\alpha=2$ is varied. All other pulse parameters are keep equal for both IR fields. 

\subsection{Numerical Results}

Figure~\ref{fig:numerical_fringes1} shows the angular distribution ($\Omega_Z$, vertical axis) of the time-averaged, macroscopic 
frequency distribution of the Poynting vector $\langle S_D\rangle(\boldsymbol{r}_d,\omega)$, defined in \eqref{eq:pointyng_freq_distrb}, as a function of the harmonic order (horizontal axis),
for two different relative CEP: $\Delta\phi = 0.0\,$ rad (panels (a) and (b)), and 
$\Delta\phi = 0.2\,$ rad.  (panels (c) and (d)), for $\Omega_X=0$. We recall that $\Delta\phi$ refers to the relative CEP between the IR fields acting on the samples denoted by $\alpha=1$ and $\alpha=2$, see Fig.~\ref{fig:scheme}. The decrease in intensity for large divergence angles
is a consequence of the macroscopic effect due to the volume averaging.
The divergence angle $\Omega_Z$ is defined in~\eqref{eq:omega_z} and depicted in Fig.~\ref{fig:scheme}.
It is to note the similarities between angular distributions of the signal at a fixed harmonic order shown 
in Fig.~\ref{fig:numerical_fringes1},  panels (b) and (d),  and the experimental data shown in Fig.~\ref{fig:FFT}.

As a function of the relative CEP phase, the overall interfering HHG signal is shifted: in panel (d), we compare the signal intensities at the harmonic $15\omega_0$
for two relative CEP, $\Delta\phi=0.0\,$ rad. and $\Delta\phi=0.2\,$ rad. The relative spectral phase of the atomic dipoles generating the radiation fields
arising from each arm are obtained from the
spectral phase of the signals shown in Fig.~\ref{fig:numerical_fringes1}. The latter
is extracted from the shift as a function of the relative CEP shown in panel (d). 

\begin{figure}[ht]%
\centering
\hspace{-0.3cm}\includegraphics[width=0.99\linewidth]{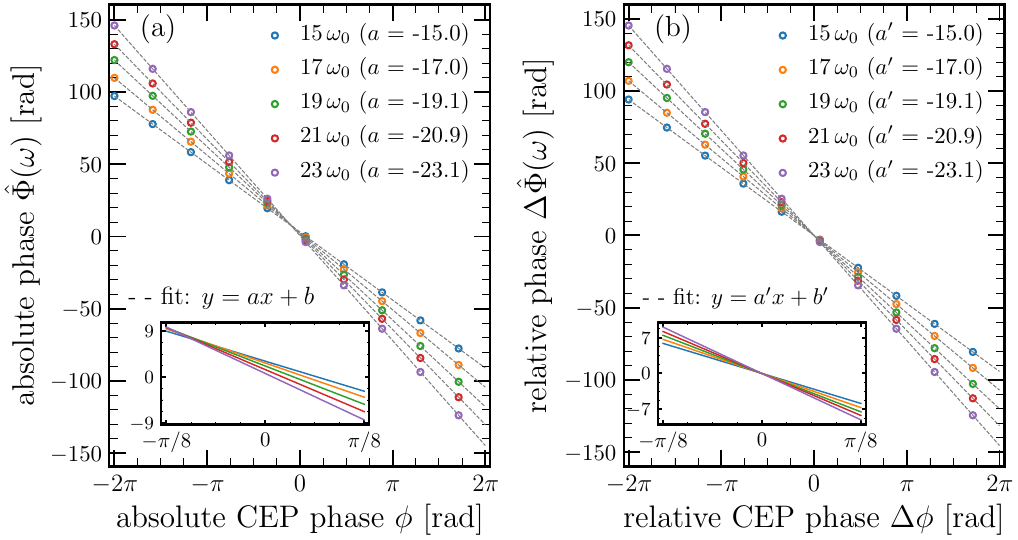}
\caption{
\textbf{Numerical Simulations. Relative Spectral Phase Retrieval}: (a) Absolute spectral phase of the dipole acceleration $\Phi_1(\omega)=\arg_\omega \mathcal{F}\left[\ddot{d}_z(t) \right](\omega)$
for different harmonic frequencies, $\omega=(2k\!+\!1)\omega_0$, as a function of the absolute CEP of the IR field in Arm 1 (in the absence of IR 2). $\Phi_1(\omega/\omega_0)$ exhibits a linear behaviour $y=ax+b$ as a function of the IR CEP with a slope $a$ that corresponds to the harmonic order. (b) Spectral phase of the interfering HHG harmonic signal arising from both arms, corresponding to the phase in the interference fringes shown in Fig.~\ref{fig:numerical_fringes1} as
a function of the relative CEP phase of the IR fields acting on the samples in arms 1 and 2. It corresponds
to the relative spectral phase $\Delta\Phi(\omega)\equiv\Phi^{(1)}_{dip}(\omega) - \Phi^{(2)}_{dip}(\omega)$ appearing in~\eqref{eq:eq_interf}. 
}
\label{fig:dp_vs_cep_supple}
\end{figure}

The quantity $\langle S_D\rangle(\boldsymbol{r}_d,\omega)$ in \eqref{eq:pointyng_freq_distrb} describes the interfering HHG radiation signal arising from both arms and depends parametrically on $\Delta\phi$, i.e.,
$\langle S_D\rangle(\boldsymbol{r}_d,\omega)=\langle S_D\rangle(\boldsymbol{r}_d,\omega;\Delta\phi)$.
To extract the phase from $\langle S_D\rangle(\boldsymbol{r}_d,\omega;\Delta\phi)$, for each $ \Delta\phi$
we introduce the transformation,
\begin{subequations}
\label{eq:myfft}
\begin{eqnarray}
\label{eq:myfft_1}
\hat{G}_{\Delta\phi}\big[\langle S_D\rangle(\boldsymbol{r}_d,\omega;\Delta\phi)\big](\omega_j) = \int dZ\,\, 
\langle S_D\rangle(\boldsymbol{r}_d,\omega;\Delta\phi)\, e^{-i\omega_j\,Z}\,,
\end{eqnarray}
with $Z\equiv\Omega_Z\,\delta z/c$ and where $\delta z\equiv z_2 - z_1$ 
is relative position of the samples along the $z-$ axis in Fig.~\ref{fig:scheme}.
As stated above, we have assumed that $x_\alpha=y_\alpha=0$ for both $\alpha=1,2$, and $n_1(\omega)=n_2(\omega)=1$, see main text.
The phase of $\langle S_D\rangle(\boldsymbol{r}_d,\omega;\Delta\phi)$ at fixed $\omega=\omega_j$ and relative CEP 
$\Delta\phi$ is then given by, 
\begin{eqnarray}
\label{eq:myfft_2}
\Delta\hat{\Phi}(\omega_j;\Delta\phi) = \arg \big \{ 
\hat{G}_{\Delta\phi}[\langle S_D\rangle(\boldsymbol{r}_d,\omega;\Delta\phi)](\omega_j)
\big\}\,.
\end{eqnarray}
\end{subequations}
Figure~\ref{fig:dp_vs_cep_supple} is repeated from the main text here for simplicity. Panel (a) shows the \emph{absolute} spectral phase of the Fourier transform of the dipole acceleration,$\hat{\ddot{d}}(\omega)$, in arm $\alpha=2$
for the harmonics $(2j+1)\omega_0$ as a function
of the \emph{absolute} CEP phase of the IR pulse acting on a single-atom in the sample $\alpha=2$. A linear fit $y=ax+b$ 
shows that, at the atomic level, the absolute spectral phase of a given high-order 
harmonic $\omega_{2k+1}=(2k+1)\omega_0$ exhibits a linear dependence on the absolute CEP phase of the IR pulse, whereby 
the slope $a$ in Fig.~\ref{fig:dp_vs_cep_supple}(a) coincides with the harmonic order $(2k+1)$. 

On the other hand, Fig.~\ref{fig:dp_vs_cep_supple}(b) shows the spectral phase of the interfering HHG radiation signal
extracted from $\langle S_D\rangle(\boldsymbol{r}_d,\omega;\Delta\phi)$ using~\eqref{eq:myfft}. This is, the phase of the 
the ``macroscopic'', time-averaged Poynting vector given by~\eqref{eq:pointyng_freq_distrb} for which
we have used the macroscopic, i.e., volume-averaged, electric fields arising from both arms $(\alpha=1,2)$ that interfere
at a common point in the detector. In Fig.~\ref{fig:dp_vs_cep_supple}(b) the horizontal axis correspond to the relative CEP
of the IR pulses acting on samples $\alpha=1$ and $\alpha=2$. It is to note that from~\eqref{eq:eq_interf}, the phase of $\langle S_D\rangle(\boldsymbol{r}_d,\omega;\Delta\phi)$ corresponds to the relative spectral phases of the atomic dipoles $\Delta\phi_{dip}(\omega) = \phi^{(1)}_{dip}(\omega) - \phi^{(2)}_{dip}(\omega)$. 
For a relative CEP $\Delta\phi=0$, we obtain a value of $\Delta\hat{\Phi}(\omega_j)=0$, see inset in Eq.~\ref{fig:numerical_fringes1}(b). For comparison, absolute phase information are shown in the inset of Fig.~\ref{fig:numerical_fringes1}(a).  
Numerical simulations are thus in good agreement with the experimental measurements reported in Fig.~(2) in the main text, showing the 
linear phase evolution of harmonics 15-21 generated in Ar as a function of fundamental phase difference.

\section{Appendix: Fundamental Theoretical model} \label{sect:Supple_Theoretical_model} 

\subsection{Microscopic Description of Laser-Induced Electromagnetic Radiation}
\label{subsec:microscopic}

To describe propagation of the incident IR/XUV fields through the samples $\alpha=1,2$ (see main text),
we solve the Maxwell equations in the Lorenz gauge. We first consider the generation of electromagnetic 
radiation due to the interaction of the incident, external radiation with a single atom. Macroscopic effects will be included as a second step, as
described in Sec.~\ref{subsec:macroscopic}. The homogeneous solution to the Maxwell equations (vanishing charge density and current density ) describe
the propagation of the incident field in the absence of the samples, i.e., free space, whereas
the particular solution to the inhomogeneous equation describe 
the radiative emission response due to the non-stationary charge distributions and 
charge currents triggered by the (total) field in the medium. Propagation of the incident field
through the medium thus results in the emission of an effective secondary radiation source that adds up to the incident field to obtain the
effective, propagated field. To obtain the particular solution, we first solve the Maxwell equations for the scalar and vector potentials, which satisfy, 
\begin{subequations}
\label{eq:lorentz_gauge}
\begin{eqnarray}
\label{eq:lorentz_gauge_Phi}
\boldsymbol{\square}\,\Phi(\boldsymbol{r},t)&=&\rho(\boldsymbol{r},t)/\epsilon_0\,,\\[0.2cm]
\label{eq:lorentz_gauge_A}
\boldsymbol{\square}\,\boldsymbol{A}(\boldsymbol{r},t)&=&\mu_0\boldsymbol{j}(\boldsymbol{r},t)\,,
\end{eqnarray} 
\end{subequations}
with $\boldsymbol{\square}\equiv\boldsymbol{\nabla}^2\, -\, \partial^2/\partial t^2$ 
and where $\rho(\boldsymbol{r},t)$ and $\boldsymbol{j}(\boldsymbol{r},t)$ denote the charge density and current density
distributions in the medium generating the secondary radiation field. Both,
$\rho(\boldsymbol{r},t)$ and $\boldsymbol{j}(\boldsymbol{r},t)$ are treated quantum mechanically. At the single-atom level, the time-dependent dipole moment is given by 
	$\boldsymbol{d}(t)=\int\boldsymbol{r}\rho(\boldsymbol{r},t)\, d^3\boldsymbol{r}$, 
with $\rho(\boldsymbol{r},t)\!=\!\langle\Psi(t)|\boldsymbol{r}\rangle\langle\boldsymbol{r}|\Psi(t)\rangle$ the charge probability density 
distribution 
and 
$|\Psi(t)\rangle$ the state vector obeying the time-dependent Schr\"odinger equation,
\begin{eqnarray}
\label{eq:schroedinger} 
	i(\partial/\partial t)\, |\Psi(t)\rangle = \hat{H}(t)|\Psi(t)\rangle\,,
\end{eqnarray}
with $\hat{H}(t)\!=\!\hat{H}_0+\hat{V}(t)$ with $\hat{H}_0$ and $\hat{V}(t)$
denoting the field-free and interaction Hamiltonians respectively.
Solutions of \eqref{eq:lorentz_gauge} obeying causality are given by the ``retarded"  scalar and vector potentials,
\begin{subequations}
\label{eq:solution_lorentz_gauge}
\begin{eqnarray}
\label{eq:solution_lorentz_gauge_Phi}
\Phi(\boldsymbol{r},t)&=&\dfrac{1}{4\pi\epsilon_0}\, \int\dfrac{\rho(\boldsymbol{r}^\prime,t_r)}{|\boldsymbol{r}-\boldsymbol{r}^\prime|}\,\, d^3\boldsymbol{r}^\prime\,,  \\[0.2cm]
\label{eq:solution_lorentz_gauge_A}
\boldsymbol{A}(\boldsymbol{r},t)&=&\dfrac{\mu_0}{4\pi}\,\int  \dfrac{\boldsymbol{j}(\boldsymbol{r}^\prime,t_r)}{|\boldsymbol{r}-\boldsymbol{r}^\prime|} d^3\boldsymbol{r}^\prime\,, 
\end{eqnarray} 
\end{subequations}
with $t_r\equiv|\boldsymbol{r}-\boldsymbol{r}^\prime|/c$ the retarded time and $c$ is the velocity of light in vacuum.  
In the far-field region, and
to the lowest order approximation in  $r/r^\prime$,
\eqref{eq:solution_lorentz_gauge} takes the asymptotic form, 
\begin{subequations}
\begin{eqnarray}
\label{eq:A_asymptotic_velocity}
	\boldsymbol{A}(\boldsymbol{r},t)&\underset{r\gg r^\prime}{=}&\dfrac{\mu_0}{4\pi r}\,\boldsymbol{\dot{d}}(t-r/c)\,.\\[0.15cm] 
\label{eq:phi2}
\Phi(\boldsymbol{r},t)&=&\dfrac{Q_e}{4\pi\epsilon_0r} + 
	\dfrac{\mu_0 c}{4\pi r}\,\, \hat{\text{\textbf{e}}}_r\cdot\boldsymbol{\dot{d}}(t-r/c)\,,\\[-0.2cm]\nonumber
\end{eqnarray}
\end{subequations}
with $Q_e \!=\!\int d^3\boldsymbol{r}^\prime\rho(\boldsymbol{r}^\prime, t^\prime)$ the total charge. In deriving \eqref{eq:A_asymptotic_velocity} 
we have assumed that the current of probability density passing through a closed surface of arbitrarily large radius from the radiation source can be neglected. Weak 
ionization probabilities are therefore implicit in the model. In spherical coordinates, the magnetic field 
$\boldsymbol{B}(\boldsymbol{r},t)=\boldsymbol{\nabla}\times\boldsymbol{A}(\boldsymbol{r},t)$ becomes 
\begin{subequations}
\label{eq:magnetic_field}
\begin{eqnarray}
B_r(\boldsymbol{r},t)&=&0\,,\\[0.3cm] 
B_\vartheta(\boldsymbol{r},t)&=&\mathcal{M}_x(r,t^\prime)\cos\varphi - \mathcal{M}_y(r,t^\prime)\sin\varphi\,,\\[0.3cm] 
B_\varphi(\boldsymbol{r},t)&=& \mathcal{M}_x(r,t^\prime)\cos\vartheta\cos\varphi -\mathcal{M}_z(r,t^\prime)\sin\vartheta
+  \mathcal{M}_y(r,t^\prime)\cos\vartheta\sin\varphi\,.
\end{eqnarray}
\end{subequations}
with  $\mathcal{M}_j(r,t^\prime)\!=\!-(\mu_0/4\pi) \big[\dot{d}_j(t\!-\!r/c)/{r^2} + \ddot{d}_j(t\!-\!r/c)/(rc)\big]$,  
and where
 $d_j$, for $j=(x,y,z)$, indicates the Cartesian components of the dipole moment.
Finally,  $\dot{d}_j(t^\prime)\equiv(\partial/\partial t)\, d(t-r/c)$ and
$\ddot{d}_j(t^\prime)\equiv(\partial^2/\partial t^2)\,d(t-r/c)$ denote the dipole velocity and acceleration, respectively. The electric field,  
$\boldsymbol{E}(\textbf{r},t)\!=\! -\boldsymbol{\nabla}\Phi(\boldsymbol{r},t)\!-\!(\partial/\partial t)\,\boldsymbol{A}(\boldsymbol{r},t)$ due to
the charge distribution and current density reads 
\begin{subequations}
\label{eq:electric_field}
\begin{eqnarray}
\label{eq:e_radial}
E_r(\boldsymbol{r},t)&=&Q_x(r,t^\prime)\sin\vartheta\cos\varphi
-Q_y(r,t^\prime)\sin\vartheta\sin\varphi +Q_z(r,t^\prime)\cos\vartheta\,,\quad\quad\\[0.4cm] 
\label{eq:e_theta}
E_\vartheta(\boldsymbol{r},t)&=& G_x(r,t^\prime)\cos\vartheta\cos\varphi
 + G_y(r,t^\prime)\cos\vartheta\sin\varphi -G_z(r,t^\prime)\sin\vartheta\,,\\[0.4cm] 
\label{eq:e_phi}
E_\varphi(\boldsymbol{r},t)&=& G_y(r,t^\prime)\cos\varphi - G_x(r,t^\prime)\sin\varphi\,,\quad\quad 
\end{eqnarray}
\end{subequations}
with 
$Q_j(r,t^\prime)=-(\mu_0 c/ 2\pi) \big[ \dot{d}_j(t^\prime)/{r^2} + d_j(t^\prime)/(r^3 c^{-1}) \big]$, and 
$G_j(r,t^\prime)=-(\mu_0c / 4\pi) \big[ \ddot{d}_j(t^\prime)/(rc) + \dot{d}_j(t^\prime)/{r^2} + d_j(t^\prime)c/{r^3}\big]$.
Note that in the far-field region, only the dipole acceleration $\ddot{\boldsymbol{d}}(t^\prime)$ contributes in \eqref{eq:magnetic_field} and~\eqref{eq:electric_field} as the
dipole and velocity counterparts fall off as $1/r^3$ and $1/r^2$, respectively. In the far-field region, the particular solutions for the radiation fields then become,
\begin{subequations}
\label{eq:Efield_final}
\begin{eqnarray}
\label{eq:Efield_final2}
E_\vartheta(\boldsymbol{r},t)&=& \dfrac{\mu_0}{4\pi r}\Big[ \ddot{d}_z(t^\prime)\sin\vartheta-\ddot{d}_y(t^\prime)\cos\vartheta\sin\varphi -\ddot{d}_x(t^\prime)\cos\vartheta\cos\varphi\, \Big]\,,\\[0.2cm] 
\label{eq:Efield_final3}
E_\varphi(\boldsymbol{r},t)&=& \dfrac{\mu_0}{4\pi r}\Big[ \ddot{d}_x(t^\prime)\sin\varphi-\ddot{d}_y(t^\prime)\cos\varphi\,\Big]\,,\\[0.2cm]
E_r(\boldsymbol{r},t)&=& 0\,,
\end{eqnarray}
\end{subequations}
for the electric radiation field at the single-atom level. Conversely, for the magnetic field, 
\begin{subequations}
\label{eq:Bfield_final}
\begin{eqnarray}
B_\vartheta(\boldsymbol{r},t)&=& \dfrac{\mu_0}{4\pi rc}\Big[ -\ddot{d}_x(t^\prime)\sin\varphi+\ddot{d}_y(t^\prime)\cos\varphi\,\Big]\\[0.2cm]
B_\varphi(\boldsymbol{r},t)&=& \dfrac{\mu_0}{4\pi rc}\Big[ \ddot{d}_z(t^\prime)\sin\vartheta-\ddot{d}_y(t^\prime)\cos\vartheta\sin\varphi-\ddot{d}_x(t^\prime)\cos\vartheta\cos\varphi\, \Big]\,,\\[0.2cm]
B_r(\boldsymbol{r},t)&=& 0\,.
\end{eqnarray}
\end{subequations}
Comparing \eqref{eq:Efield_final} and ~\eqref{eq:Bfield_final}, 
$B_\vartheta(\textbf{r},t)=-E_\varphi(\boldsymbol{r},t)/c$, and 
$B_\varphi(\textbf{r},t)  =+E_\vartheta(\boldsymbol{r},t)/c$. In the far-field limit, the corresponding 
Poynting vector $\boldsymbol{S}(\boldsymbol{r},t)\!=\!\boldsymbol{E}(\boldsymbol{r},t)\times\boldsymbol{B}(\boldsymbol{r},t)/\mu_0$ 
measuring the energy-flux density per unit area per unit time, reduces to
\begin{eqnarray}
\label{eq:Poynting1} 
\boldsymbol{S}(\boldsymbol{r},t)= \sqrt{\dfrac{\epsilon_0}{\mu_0}} \, \big[ |E_\vartheta(\boldsymbol{r},t)|^2 + |E_\varphi(\boldsymbol{r},t)|^2 \big]\,\hat{\text{\textbf{e}}}_r\,.  
\end{eqnarray}

\subsection{Macroscopic Propagation and Volume Averaging}
\label{subsec:macroscopic}

Making use of the property $\mathcal{F}\{d(t-r/c)\}(\omega)\!=\!e^{i\omega r/c}\mathcal{F}\{d(t)\}(\omega)$,
the frequency distribution of the electric field component defined by the first term in the RHS in~\eqref{eq:Efield_final2}  
at a point  $\boldsymbol{r}_d=(x_d,y_d,z_d)$ on the detector plane
due to the atomic charge distribution
located at the position $\boldsymbol{r}_i=(x_i,y_i,z_i)$  is given by 
\begin{eqnarray}
\label{eq:micro_final}
E_{\vartheta}(\boldsymbol{r}_d-\boldsymbol{r}_i,\omega) = \dfrac{\mu_0}{4\pi |\boldsymbol{r}_d-\boldsymbol{r}_i |}\,
\sin{\vartheta_{d,i}}\,\, e^{i\omega |\boldsymbol{r}_d-\boldsymbol{r}_i|/c}\, \hat{\ddot{d}}_z(\omega;[E_{ext}(\boldsymbol{r}_i)])\,.  
\end{eqnarray}
The position vectors $\boldsymbol{r}_d$ and $\boldsymbol{r}_i$ are defined with respect to 
the coordinate system $\mathcal{R}=(x,y,z)$ shown in Fig.~\ref{fig:scheme} whilst $\theta_{d,i}$
denotes the spherical polar angle defined by an arbitrary $\tilde{z}-$ axis in the atomic frame of reference centered at $\boldsymbol{r}_i$ 
and the vector $\boldsymbol{r}_d-\boldsymbol{r}_i$.  
The quantity $\hat{\ddot{d}}_z(\omega;[E_{ext}(\boldsymbol{r}_i)])$ in~\eqref{eq:micro_final} stands for the Fourier Transform of the 
atomic dipole acceleration obtained by solving the time-dependent Schr\"odinger equation for the atomic system
located at $\boldsymbol{r}_i$ in $\mathcal{R}$ subject to the electric field $E_{ext}(t,\boldsymbol{r}_i)$. The latter may or may not
be a function of the position $\boldsymbol{r}_i$ within the  samples $\alpha=1,2$ macroscopically characterized by the
refractive index $n_\alpha(\omega)$ in Fig.~\ref{fig:scheme}. Analytical models allowing to obtain closed-form expressions 
for HHG radiation fields that incorporates  macroscopic effects have 
been obtained by solving the 1D-Maxwell equations in a 
infinitely thin, infinitely long gas distribution~\cite{Baggesen2011} as well as by accounting for 
non-uniform intensity distribution of focused Gaussian beams in the slowly-varying envelope approximation~\cite{Shore1989}.  Here, we resort to define 
the ``net'' electric field originating from the sample $\alpha$ as net 
contribution of the individual atomic charge distributions within each sample, namely, 
\begin{eqnarray}
\label{eq:sum_ei}
E^{(\alpha)}_{\vartheta}(\boldsymbol{r}_d, \omega) = \sum^{N_{\alpha}}_{i=1} E_{\vartheta}(\boldsymbol{r}_d-\boldsymbol{r}_i,\omega)\,,
\end{eqnarray}
with $N_\alpha$ the number of microscopic charge distributions (atoms) in the sample $\alpha$ and 
$E_{\vartheta}(\boldsymbol{r}_d-\boldsymbol{r}_i,\omega)$ defined in~\eqref{eq:micro_final}. We approximate~\eqref{eq:sum_ei} 
according to
\begin{eqnarray}
\label{eq:continuum_limit}
E^{(\alpha)}_{\vartheta}(\boldsymbol{r}_d, \omega) = 
\int_{V_{\alpha}} d^3\boldsymbol{r}^\prime\,\,
\rho_{\alpha}(\boldsymbol{r}^\prime)\,\,E_{\vartheta}(\boldsymbol{r}_d-\boldsymbol{r}^\prime,\omega)
\end{eqnarray}
with $\rho_{\alpha}(r^\prime)$ the macroscopic density of atoms
in the sample $\alpha$. Using~\eqref{eq:micro_final},~\eqref{eq:continuum_limit} becomes, 
\begin{eqnarray}
\label{eq:E_end}
E^{(\alpha)}_{\vartheta}(\boldsymbol{r}_d, \omega) &=& \dfrac{\mu_0}{4\pi} 
\int_{V_{\alpha}} d^3\boldsymbol{r}^\prime\,\,
\dfrac{e^{i\frac{\omega}{c}|\boldsymbol{r}_d - \boldsymbol{r}^\prime|}}{\boldsymbol{r}_d-\boldsymbol{r}^\prime}\sin\vartheta^\prime_{d}\,\, \rho_{\alpha}(\boldsymbol{r}^\prime)\,\,
\hat{\ddot{d}}_z(\omega;[E_{ext}(\boldsymbol{r}^\prime)])\,.
\end{eqnarray}
In the coordinate system $\mathcal{R}$, the volume integration is given, according to Fig.~\ref{fig:scheme}, by,
\begin{eqnarray}
\label{eq:int_equiv}
\int_{V_{\alpha}} d^3\boldsymbol{r}^\prime \equiv 
\int^{{y_{\alpha}+\frac{\Delta y}{2}}}_{y_{\alpha}-\frac{\Delta y}{2}}\,dy^\prime 
\int^{{x_{\alpha}+\frac{\Delta x}{2}}}_{y_{\alpha}-\frac{\Delta x}{2}}dx^\prime
\int^{{z_{\alpha}+\frac{\Delta z}{2}}}_{y_{\alpha}-\frac{\Delta z}{2}}dz^\prime\,,
\end{eqnarray}
where $\Delta{x}$, $\Delta{y}$ and $\Delta{z}$ denote the dimensions of the samples in arms $\alpha=1,2$, 
and $\boldsymbol{r}_\alpha=(x_\alpha,y_\alpha,z_\alpha)$ the center-of-mass position of the samples 
represented by the yellow ($\alpha=1$) and red ($\alpha=2$) bullets in Fig.~\ref{fig:scheme}.
In our numerical simulations, we have set $\boldsymbol{r}_{\alpha} = (0,0,z_\alpha)$ 
with $z_2=-z_1$, see vectors $\boldsymbol{r}_1$ and $\boldsymbol{r}_2$ in Fig.~\ref{fig:scheme}. 

Close-form expressions to~\eqref{eq:continuum_limit} can be obtained by considering
an homogeneous gas density,  
with $N_\alpha$ atoms per unit volume $V_\alpha$ in the sample $\alpha$,
i.e., $\rho_{\alpha}(\boldsymbol{r}^\prime)=N_\alpha/V_\alpha$.
As a second approximation, we consider a plane-wave distribution 
for the incident external IR field  $E_{ext}(t,\boldsymbol{r}^\prime)$, 
\begin{eqnarray}
\label{eq:E_ext}
\boldsymbol{E}_{ext}(t,\boldsymbol{r}^\prime) = \int d\omega\,\, \hat{E}_{ext}(\omega)\, e^{-i(\omega t - k(\omega)\, y^\prime)}\,\, \hat{e}_z \,, 
\end{eqnarray}
see Fig.~\ref{fig:scheme} for choice of coordinates, with constant spatial distribution
along the propagation direction within the width of the sample. With these approximations, we may neglect 
the spatial dependencies of $\rho_a(\boldsymbol{r})\approx N_\alpha/V_\alpha$ and that of the dipole acceleration which
depends implicitly on $E_{ext}(t,\boldsymbol{r}^\prime)$, i.e., $\hat{\ddot{d}}(\omega;[E_{ext}(\boldsymbol{r}^\prime)])\approx\hat{\ddot{d}}(\omega)$. 
Finally, in the fair-field limit, the distance between
a point $\boldsymbol{r}_d$ in the detection plane and a point $\boldsymbol{r}^\prime$ in the sample can be approximated according to,
\begin{eqnarray}
|\boldsymbol{r}_d-\boldsymbol{r}^\prime| &=& (y_d-y^\prime)\sqrt{ 1+\dfrac{(x_d-x^\prime)^2}{(y_d-y^\prime)^2} + \dfrac{(x_d-x^\prime)^2}{(y_d-y^\prime)^2}}\nonumber\\[0.2cm]
&\approx& (y_d-y^\prime) + \dfrac{1}{2} \dfrac{(x_d-x^\prime)^2}{(y_d-y^\prime)} +  \dfrac{1}{2} \dfrac{(z_d-z^\prime)^2}{(y_d-y^\prime)} + \dots 
\end{eqnarray}
Retaining the zeroth order approximation for the denominator $1/|\boldsymbol{r}_d - \boldsymbol{r}^\prime|$ in~\eqref{eq:E_end} and first order approximation in
the exponents $\exp(iw|\boldsymbol{r}-\boldsymbol{r}^\prime|/c)$, we obtain
\begin{eqnarray}
\label{eq:int3}
\small
E^{(\alpha)}_{\vartheta}(\boldsymbol{r}_d, \omega) =\dfrac{\mu_0}{4\pi}\left(\dfrac{N_{\alpha}}{V_\alpha}\right)\,\, \hat{\ddot{d}}_z(\omega) 
\int^{{y_{\alpha}+\frac{\Delta y}{2}}}_{y_{\alpha}-\frac{\Delta y}{2}}\,dy^\prime 
\dfrac{e^{i\frac{\omega}{c}(y_d-y^\prime)}}{y_d-y^\prime}
\int^{{x_{\alpha}+\Delta x/2}}_{y_{\alpha}-\Delta x/2}dx^\prime
e^{-i\frac{\omega}{c}\frac{x_d}{y_d}\,x^\prime}
\int^{{z_{\alpha}+\frac{\Delta z}{2}}}_{y_{\alpha}-\Delta z/2}dz^\prime\,\,
e^{-i\frac{\omega}{c}\frac{z_d}{y_d}\,z^\prime}
\end{eqnarray}
For small divergence angles $\Omega_Z$ and $\Omega_X$  
depicted by the  blue, resp. red lines in Fig.~\ref{fig:scheme},  we approximate both quantities according to its
first order expansion 
\begin{eqnarray}
\label{eq:omega_x}
\dfrac{x_d}{y_d} &\equiv& \tan(\Omega_X) \approx \Omega_X\\[0.2cm]
\label{eq:omega_z}
\dfrac{z_d}{y_d} &\equiv& \tan(\Omega_Z) \approx \Omega_Z\,.
\end{eqnarray}
With these approximations, analytical integration of~\eqref{eq:int3} finally gives 
\begin{eqnarray}
\label{eq:E_macro}
E^{(\alpha)}_{\vartheta}(\boldsymbol{r}_d, \omega) &=&\dfrac{\mu_0}{4\pi\,|y_d|}\, N_{\alpha}\,\, \hat{\ddot{d}}_z(\omega)\,\, 
e^{i\frac{\omega}{c}(y_d-y_{\alpha})}\, e^{-i\frac{\omega}{c}\,x_\alpha\,\Omega_X }\, e^{-i\frac{\omega}{c}\,z_\alpha\,\Omega_Z }\nonumber\\[0.2cm] 
&\times&\mathrm{sinc}\left(\dfrac{\omega}{c}\dfrac{\Delta y}{2}\right)\,
\mathrm{sinc}\left(\frac{\omega}{c}\frac{\Delta x}{2}\Omega_X  \right)\,
\mathrm{sinc}\left(\frac{\omega}{c}\frac{\Delta z}{2} \Omega_Z \right)\,,
\end{eqnarray}
where $E^{(\alpha)}_{\vartheta}(\boldsymbol{r}_d, \omega)\equiv E^{(\alpha)}_{\vartheta}(\boldsymbol{r}_d-\boldsymbol{r}_a,\omega)$ is a function of the center-of-mass position $\boldsymbol{r}_a\equiv(x_a,y_a,z_a)$ of the sample $\alpha=1,2$. 
After integrating~\eqref{eq:int3}, the volume $V_\alpha$ in the denominator is compensated by the same quantity 
that eventually appears in the numerator. Integration over a continuum and homogeneous ensemble of atomic dipoles  
subject to an homogeneous external field $E_{ext}(t)$
leads to a sharply focused sinc-like spatial electric field distribution 
as opposed to its single-atom counterpart, i.e., compare~\eqref{eq:E_macro}
and first term in the RHS of~\eqref{eq:Efield_final2}. Finally, it is straightforward to show that 
the above procedure to obtain the closed-form expression for the macroscopic field in~\eqref{eq:E_macro} is totally equivalent to solving the well-known inhomogeneous equation in the radiation zone for the macroscopic field~\cite{Shore1989},
\begin{eqnarray}
\label{eq:maxwell_macro}
\left(\boldsymbol{\nabla}^2 -\dfrac{\partial^2}{\partial t^2} \right)\boldsymbol{E}(\boldsymbol{r},t)= \mu_0\dfrac{\partial^2}{\partial t^2}\boldsymbol{P}(t)\,,
\end{eqnarray}
for a homogeneous ensemble of $N$ atoms per unit volume $V$,
with $\boldsymbol{P}(t)=N_{\alpha}\, \langle\boldsymbol{d}^{(\alpha)}(t)\rangle_{V}$ the macroscopic polarization of the sample $\alpha$ defined by the dipole density $\langle\boldsymbol{d}^{(\alpha)}(t)\rangle_{V}=(1/V)\int d^3\boldsymbol{r}_i\, \boldsymbol{d}(t;\boldsymbol{r}_i)$, with $\boldsymbol{d}(t;\boldsymbol{r}_i)$ the single-atom dipole moment with center-of-mass location $\boldsymbol{r}_i$.

\section{Appendix: Superposition of OAM} \label{sect:Supple_OAM_model}

\section{Superposition of LG modes carrying opposite OAM values}
\subsection{LG mode's description}
The complex field amplitude of a Laguerre-Gaussian (LG) mode can be written as (in cylindrical coordinates):
\begin{eqnarray}
    E(r,\theta,z)= E_{0} \frac{w_{0}}{w(z)} \mathrm{e}^{-\frac{r^2}{w(z)^2}} \left(\frac{\sqrt{2}r}{w(z)}\right)^{|l|} \mathrm{e}^{-\frac{ikr^2}{2 R(z)}} \mathrm{e}^{i\Phi_{G}} \mathrm{e}^{il\theta},
    \label{eqn1}
\end{eqnarray}
where we have used a zero-radial index i.e., $p=0$, to ensure that there's a single bright ring in the transverse cross-section of the beam. We define the beam parameters as follows:
\begin{eqnarray}
    w(z)&=&w_{0}\sqrt{1+\left(\frac{z}{z_{R}}\right)^2} \nonumber \\
    R(z)&=& z \left(1+\frac{z_{R}^2}{z^2}\right) \nonumber \\
    \Phi_{G}&=&-(|l|+1)\arctan\left(\frac{z}{z_{R}}\right) \nonumber \\
    z_{R}&=& \frac{1}{2}kw_{0}^2.
    \label{eqn2}
\end{eqnarray}
Here, $w(z),R(z),\Phi_{G}$, and $z_{R}$ represent the beam width at some finite propagation distance $z$, the radius of curvature of the wavefront at $z$, the Gouy phase at $z$, and the Rayleigh range of the beam, respectively. Furthermore, $E_{0},w_{0}$, and $l$ denote the peak amplitude, the beam waist size, and the topological charge (TC) (alternatively, an OAM of $l\hbar$ per photon) of the LG beam, respectively.
\subsection{Superposition of LG modes with OAMs $+1$ and $-1$}
If we consider the plane just after the spatial light modulator (SLM) as the source plane for the generation of superposed LG modes, the complex field amplitude of the superposed LG modes can be expressed as:
\begin{eqnarray}
    E_{total}(r,\theta) &=& E_{1} r \mathrm{e}^{-\frac{r^2}{w_{0}^2}}\left[\mathrm{e}^{i\theta}+\mathrm{e}^{-i\theta}\right] \nonumber \\
    &=& 2 E_{1} r \mathrm{e}^{-\frac{r^2}{w_{0}^2}} \cos(\theta),
    \label{eqn3}
\end{eqnarray}
where we have used $l=+1$ and $-1$ for the two LG modes and $E_{1}=\frac{\sqrt{2}}{w_{0}}E_{0}$. Here, ($r,\theta$) are the coordinates of the resultant beam at the source plane.
Utilizing Eq.~(\ref{eqn3}), the intensity of the resultant beam can be calculated as:
\begin{eqnarray}
    I_{total}(r,\theta)&=&|E_{total}(r,\theta)|^2 \nonumber \\
    &=& 4 |E_{1}|^2 r^2 \mathrm{e}^{-2\frac{r^2}{w_{0}^2}} \cos^{2}(\theta) \nonumber \\
    &=& 2 |E_{1}|^2 r^2 \mathrm{e}^{-2\frac{r^2}{w_{0}^2}} \left[1+\cos(2\theta)\right].
    \label{eqn4}
\end{eqnarray}

\begin{figure}[h!]
\centering
\includegraphics[width=\textwidth]{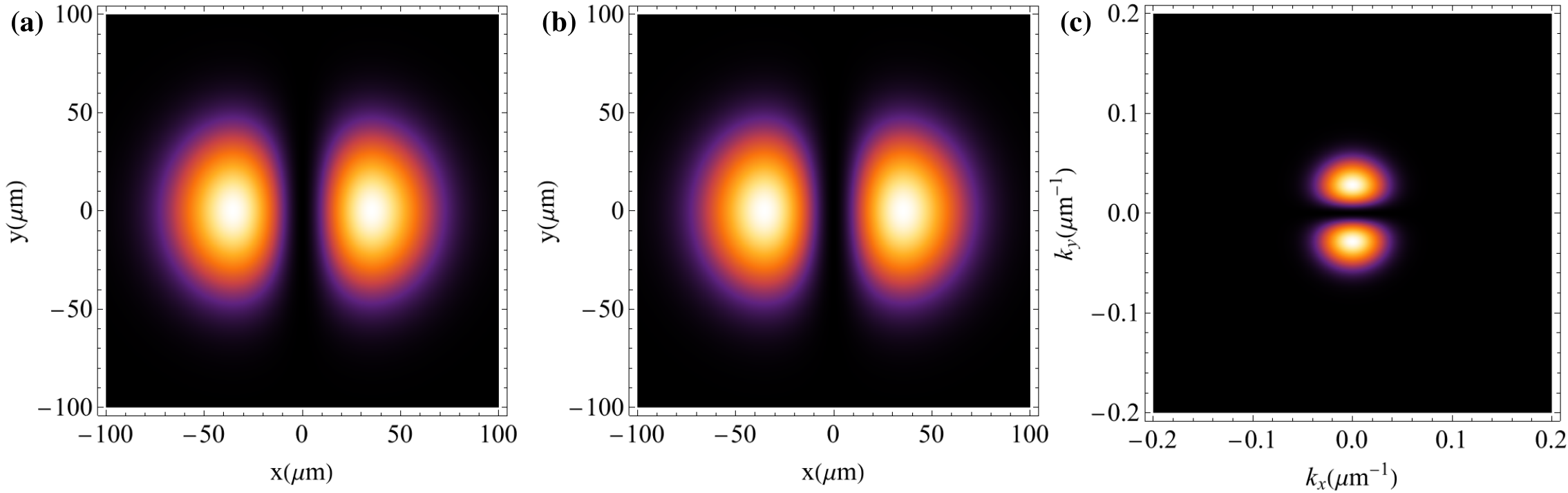}
\caption{
\textbf{(a)} Superposition of LG modes with OAM $l=+1$ and $l=-1$.
\textbf{(b)} Transverse intensity distribution of an $HG_{1,0}$ mode. 
\textbf{(c)} Superposed LG modes at the focal plane of the lens.
}
\label{Fig_OAM_Theory}
\end{figure}
It can be seen from Fig.~\ref{Fig_OAM_Theory}(a) that the transverse intensity distribution of the resultant beam shows lobed structure i.e., two bright lobes along the horizontal direction (labeled as $x$-axis) with an intensity null (or, node) at $x=0$. This intensity pattern resembles with that of the $HG_{1,0}$ mode and the resultant structure does not carry a net OAM.

To better understand the intensity pattern shown in Fig.~\ref{Fig_OAM_Theory}(a), let's revisit some fundamental concepts in paraxial optics. If one solves the paraxial Helmholtz equation in Cartesian coordinates, one typically obtains $HG$ modes as the solution to the paraxial Helmholtz equation with rectangular symmetry. At the source plane ($z=0$), the complex field amplitude of an $HG$ mode can be written as $HG_{m,n}\propto H_{m}\left(\frac{\sqrt{2}x}{w_{0}}\right) H_{n}\left(\frac{\sqrt{2}y}{w_{0}}\right) \mathrm{e}^{-\frac{x^2+y^2}{w_{0}^2}}$, where, $H_{m}(..)$ and $H_{n}(..)$ are Hermite polynomials of order $m$ and $n$, respectively, and ($x,y$) denote the Cartesian coordinates of the beam. Therefore, for the $HG_{1,0}$ mode, the field amplitude becomes $HG_{1,0}\propto x \mathrm{e}^{-\frac{x^2+y^2}{w_{0}^2}}$ i.e., antisymmetric in $x$ (remember that $H_{0}(x)=1$ and $H_{1}(x)=2x$). It can be found that, for $x=0$, the amplitude vanishes and, therefore, we see null intensity between the lobes. It is important to highlight that this intensity null is not related to the presence of a phase singularity in the beam profile, rather it is connected to the presence of a phase discontinuity at $x=0$ and a phase shift of $\pi$ between the intensity lobes. In other words, this null intensity can be connected to the destructive interference at $x=0$ due to equal and opposite contributions from both the lobes. However, the phase is uniform within each lobe. In Fig.~\ref{Fig_OAM_Theory}(b), we plot the transverse intensity distribution for the $HG_{1,0}$ mode. A comparison between Figs.~\ref{Fig_OAM_Theory} (a) and (b) reveals that both these intensity patterns are precisely the same. Thus, this behaviour leads to the following conclusion: an $HG$ mode can always be written as the superposition of LG modes and the orientation of the lobes depend on the phase difference between the two superposed modes. This absolutely makes sense because LG modes present a family of orthogonal modes and any light field can be represented as a superposition of LG modes. Similar intuition can be drawn for the HG modes as well, because they also form an orthonormal basis i.e., any light field can be represented as a superposition of HG modes.

For instance, if we change the sign between the terms $\mathrm{e}^{i\theta}$ and $\mathrm{e}^{-i\theta}$ from $'+'$ to $'-'$ in Eq.~(\ref{eqn3}), we will end up with intensity lobes along the vertical direction (labeled as $y$-axis) which resembles to that of the $HG_{0,1}$ mode.

We further use the azimuthal Fourier transform of the field in Eq.~(\ref{eqn3}) to analyze its OAM spectrum. The azimuthal Fourier transform of a beam $E(r,\theta)$ is given by $E(r,\theta)=\sum_{l=-\infty}^{\infty}c_{l}(r)\mathrm{e}^{il\theta}$, where $c_{l}(r)$ denotes the OAM spectrum at radius r. Therefore, $c_{l}(r)=\frac{1}{2\pi}\int_{0}^{2\pi}E(r,\theta)\mathrm{e}^{-il\theta}d\theta$. If we plug Eq.~(\ref{eqn3}) in the place for $E(r,\theta)$, we find that $c_{1}(r)=c_{-1}(r)=2E_{1}r\mathrm{e}^{-\frac{r^2}{w_{0}^2}}$. For any other values of $l$ (apart from $l=+1$ and $l=-1$), $c_{l}=0$. This confirms that we have a coherent superposition of two LG modes with $l=+1$ and $l=-1$ in the field described by Eq.~(\ref{eqn3}) (also in the $HG_{1,0}$ mode).

\subsection{Focusing of the superposed LG modes}
Now, we want to see the structure of the beam at the focal plane after focusing through an optical lens. An optical lens typically performs the Fourier transformation on the input light field (known from Fourier optics). Quantitatively, this can explained using the Hankel transform. For a function that has both radial and angular dependence i.e., $f=f(r,\theta)$, Bessel functions of higher-order can be utilized to perform the 2D-Fourier transform of the input light field. For instance, if $f(r,\theta)=f(r) \cos(\theta)$, then the 2D-Fourier transform can be written as:
\begin{eqnarray}
    F(\kappa,\varphi)=2\pi \sin(\varphi) \int_{0}^{\infty}f(r) J_{1}(\kappa r) rdr
    \label{eqn5}
\end{eqnarray}
If we closely look at Eq.~(\ref{eqn3}), the complex field amplitude has the form of $f(r) \cos(\theta)$ precisely, where $f(r)=2E_{1}r\mathrm{e}^{-\frac{r^2}{w_{0}^2}}$. If we plug the expression for $f(r)$ in Eq.~(\ref{eqn5}) and write down the transformed electric field distribution in the Fourier space, we get:
\begin{eqnarray}
    \tilde{E}(\kappa,\varphi)&=&4\pi E_{1} \sin(\varphi) \int_{0}^{\infty}r^2 \mathrm{e}^{-\frac{r^2}{w_{0}^2}} J_{1}(\kappa r)dr \nonumber \\
    &=& 4\pi E_{1} \sin(\varphi) \frac{\kappa}{4*\left(\frac{1}{w_{0}^4}\right)} \left._{1}F_{1}\right.\left(2;2;-\frac{\kappa^2 w_{0}^2}{4}\right)
    \label{eqn6}
\end{eqnarray}
where $\left._{1}F_{1}\right.(..;..;..)$ is the Kummer confluent hypergeometric function and $(\kappa,\varphi)$ are the Fourier space coordinates. Utilizing Eq.~(\ref{eqn6}), we plot the intensity distribution of the superposed LG modes in the Fourier space. The result is shown in Fig.~\ref{Fig_OAM_Theory}(c).
It can be seen from the figure that the lobes are oriented vertically. However, in Fig.~\ref{Fig_OAM_Theory}(a), the lobes were oriented horizontally before the lens. This result is not surprising at all as this is what the Fourier transform precisely does. Furthermore, similar intuitions 
can be drawn about the local phase variations and the net OAM content of this structure.

\bibliography{bibfile_refs}

@article{breitInterpretationDiracsTheory1928,
  title = {An {{Interpretation}} of {{Dirac}}'s {{Theory}} of the {{Electron}}},
  author = {Breit, G.},
  year = 1928,
  month = jul,
  journal = {Proceedings of the National Academy of Sciences of the United States of America},
  volume = {14},
  number = {7},
  pages = {553--559},
  issn = {0027-8424},
  doi = {10.1073/pnas.14.7.553},
  urldate = {2024-03-15},
  pmcid = {PMC1085609},
  pmid = {16587362}
}

@book{greinerRelativisticQuantumMechanics1995,
  title = {Relativistic {{Quantum Mechanics}}},
  author = {Greiner, Walter},
  year = 1995,
  publisher = {Springer},
  address = {Berlin, Heidelberg},
  doi = {10.1007/978-3-642-88082-7},
  urldate = {2024-03-15},
  isbn = {978-3-540-99535-7 978-3-642-88082-7},
  langid = {english}
}

@book{schrodingerFreeMovementRelativistic1930,
  title = {On the Free Movement in Relativistic Quantum Mechanics},
  author = {Schrodinger, E},
  year = 1930,
  number = {881393652},
  publisher = {www.worldcat.org/oclc/}
}

@article{romanZitterbewegungDiracElectron2003,
  title = {The {{Zitterbewegung}} for a {{Dirac}} Electron Driven by an Intense Laser Field},
  author = {Roman, Julio San and Roso, Luis and Plaja, Luis},
  year = 2003,
  month = may,
  journal = {Journal of Physics B: Atomic, Molecular and Optical Physics},
  volume = {36},
  number = {11},
  pages = {2253},
  issn = {0953-4075},
  doi = {10.1088/0953-4075/36/11/310},
  urldate = {2024-03-15},
  langid = {english}
}

@article{rusinZitterbewegungElectronsGraphene2008,
  title = {Zitterbewegung of Electrons in Graphene in a Magnetic Field},
  author = {Rusin, Tomasz M. and Zawadzki, Wlodek},
  year = 2008,
  month = sep,
  journal = {Physical Review B},
  volume = {78},
  number = {12},
  pages = {125419},
  publisher = {American Physical Society},
  doi = {10.1103/PhysRevB.78.125419},
  urldate = {2024-10-25}
}

@article{luanZitterbewegungNewDirac2018,
  title = {Zitterbewegung near New {{Dirac}} Points in Graphene Superlattices},
  author = {Luan, Jianli and Li, Shangyang and Ma, Tianxing and Wang, Li-Gang},
  year = 2018,
  month = sep,
  journal = {Journal of Physics: Condensed Matter},
  volume = {30},
  number = {39},
  pages = {395502},
  publisher = {IOP Publishing},
  issn = {0953-8984},
  doi = {10.1088/1361-648X/aadbe0},
  urldate = {2024-10-25},
  langid = {english}
}

@article{lovettObservationZitterbewegungPhotonic2023,
  title = {Observation of {{Zitterbewegung}} in Photonic Microcavities},
  author = {Lovett, Seth and Walker, Paul M. and Osipov, Alexey and Yulin, Alexey and Naik, Pooja Uday and Whittaker, Charles E. and Shelykh, Ivan A. and Skolnick, Maurice S. and Krizhanovskii, Dmitry N.},
  year = 2023,
  month = may,
  journal = {Light: Science \& Applications},
  volume = {12},
  number = {1},
  pages = {126},
  publisher = {Nature Publishing Group},
  issn = {2047-7538},
  doi = {10.1038/s41377-023-01162-x},
  urldate = {2024-10-25},
  copyright = {2023 The Author(s)},
  langid = {english}
}

@article{gouanereExperimentalObservationCompatible2008,
  title = {Experimental Observation Compatible with the Particle Internal Clock in a Channeling Experiment},
  author = {Gouan{\`e}re, M. and Spighel, M. and Cue, N. and Gaillard, M.J. and Genre, R. and Kirsch, R. and Poizat, J.C. and Remillieux, J. and Catillon, P. and Roussel, L.},
  year = 2008,
  journal = {Annales de la Fondation Louis de Broglie},
  volume = {33},
  number = {1-2},
  pages = {85--91}
}

@article{remillieuxHighEnergyChannelling2015,
  title = {High Energy Channelling and the Experimental Search for the Internal Clock Predicted by {{Louis}} de {{Broglie}}},
  author = {Remillieux, J. and Artru, X. and Bajard, M. and Chehab, R. and Chevallier, M. and Curceanu, C. and Dabagov, S. and Dauvergne, D. and Gu{\'e}rin, H. and Gouan{\`e}re, M. and Kirsch, R. and Krimmer, J. and Poizat, J. -C. and Ray, C. and Takabayashi, Y. and Testa, E.},
  year = 2015,
  month = jul,
  journal = {Nuclear Instruments and Methods in Physics Research Section B: Beam Interactions with Materials and Atoms},
  series = {Proceedings of the 6th {{International Conference Channeling}} 2014:``{{Charged}} \& {{Neutral Particles Channeling Phenomena}}'' {{October}} 5-10, 2014, {{Capri}}, {{Italy}}},
  volume = {355},
  pages = {193--197},
  issn = {0168-583X},
  doi = {10.1016/j.nimb.2015.02.005},
  urldate = {2024-10-25}
}

@article{sennaryAttosecondQuantumUncertainty2025,
  title = {Attosecond Quantum Uncertainty Dynamics and Ultrafast Squeezed Light for Quantum Communication},
  author = {Sennary, Mohamed and {Rivera-Dean}, Javier and ElKabbash, Mohamed and Pervak, Vladimir and Lewenstein, Maciej and Hassan, Mohammed Th},
  year = 2025,
  month = oct,
  journal = {Light: Science \& Applications},
  volume = {14},
  number = {1},
  pages = {350},
  publisher = {Nature Publishing Group},
  issn = {2047-7538},
  doi = {10.1038/s41377-025-02055-x},
  urldate = {2025-11-05},
  copyright = {2025 The Author(s)},
  langid = {english}
}

@article{tzurGenerationSqueezedHighorder2024,
  title = {Generation of Squeezed High-Order Harmonics},
  author = {Tzur, Matan Even},
  year = 2024,
  journal = {Physical Review Research},
  volume = {6},
  number = {3},
  doi = {10.1103/PhysRevResearch.6.033079}
}

@article{de-la-penaQuantumElectrodynamicsHighHarmonic2025,
  title = {Quantum {{Electrodynamics}} in {{High-Harmonic Generation}}: {{Multitrajectory Ehrenfest}} and {{Exact Quantum Analysis}}},
  shorttitle = {Quantum {{Electrodynamics}} in {{High-Harmonic Generation}}},
  author = {{de-la-Pe{\~n}a}, Sebasti{\'a}n and Neufeld, Ofer and Even Tzur, Matan and Cohen, Oren and Appel, Heiko and Rubio, Angel},
  year = 2025,
  month = jan,
  journal = {Journal of Chemical Theory and Computation},
  volume = {21},
  number = {1},
  pages = {283--290},
  publisher = {American Chemical Society},
  issn = {1549-9618},
  doi = {10.1021/acs.jctc.4c01206},
  urldate = {2025-11-05}
}

@article{moiseyevConditionsAnalogQED2023,
  title = {The Conditions for the Analog of {{QED}} Photons in Semi-Classical Periodically Driven Systems},
  author = {Moiseyev, Nimrod and Even Tzur, Matan},
  year = 2023,
  month = dec,
  journal = {Journal of Optics},
  volume = {26},
  number = {2},
  pages = {025501},
  publisher = {IOP Publishing},
  issn = {2040-8986},
  doi = {10.1088/2040-8986/ad15eb},
  urldate = {2025-11-05},
  langid = {english}
}

@article{gorlachHighharmonicGenerationDriven2023,
  title = {High-Harmonic Generation Driven by Quantum Light},
  author = {Gorlach, Alexey and Tzur, Matan Even and Birk, Michael and Kr{\"u}ger, Michael and Rivera, Nicholas and Cohen, Oren and Kaminer, Ido},
  year = 2023,
  month = nov,
  journal = {Nature Physics},
  volume = {19},
  number = {11},
  pages = {1689--1696},
  publisher = {Nature Publishing Group},
  issn = {1745-2481},
  doi = {10.1038/s41567-023-02127-y},
  urldate = {2025-11-05},
  copyright = {2023 The Author(s), under exclusive licence to Springer Nature Limited},
  langid = {english}
}

@misc{tzurMeasuringControllingBirth2025a,
  title = {Measuring and Controlling the Birth of Quantum Attosecond Pulses},
  author = {Tzur, Matan Even and Mor, Chen and Yaffe, Noa and Birk, Michael and Rasputnyi, Andrei and Kneller, Omer and Nisim, Ido and Kaminer, Ido and Kr{\"u}ger, Michael and Dudovich, Nirit and Cohen, Oren},
  year = 2025,
  month = feb,
  number = {arXiv:2502.09427},
  eprint = {2502.09427},
  primaryclass = {physics},
  publisher = {arXiv},
  doi = {10.48550/arXiv.2502.09427},
  urldate = {2025-11-05},
  archiveprefix = {arXiv}
}

@article{tross_interferometer,
  title={Self referencing attosecond interferometer with zeptosecond precision},
  author={Tross, Jan and Kolliopoulos, Georgios and Trallero-Herrero, Carlos A},
  journal={Optics Express},
  volume={27},
  number={16},
  pages={22960--22969},
  year={2019},
  publisher={Optical Society of America}
}

@article{camper_high_2019,
	title = {High relative-phase precision beam duplicator for mid-infrared femtosecond pulses},
	volume = {44},
	copyright = {© 2019 Optical Society of America},
	issn = {1539-4794},
	url = {https://opg.optica.org/ol/abstract.cfm?uri=ol-44-22-5465},
	doi = {10.1364/OL.44.005465},
	number = {22},
	urldate = {2024-02-27},
	journal = {Optics Letters},
	author = {Camper, Antoine and Park, Hyunwook and Hageman, Stephen J. and Smith, Greg and Auguste, Thierry and Agostini, Pierre and DiMauro, Louis F.},
	year = {2019},
	pages = {5465--5468},
}

@article{hadrich2015exploring,
  title={Exploring new avenues in high repetition rate table-top coherent extreme ultraviolet sources},
  author={H{\"a}drich, Steffen and Krebs, Manuel and Hoffmann, Armin and Klenke, Arno and Rothhardt, Jan and Limpert, Jens and T{\"u}nnermann, Andreas},
  journal={Light: Science \& Applications},
  volume={4},
  number={8},
  pages={e320--e320},
  year={2015},
  publisher={Nature Publishing Group}
}

@article{wang2015bright,
  title={Bright high-repetition-rate source of narrowband extreme-ultraviolet harmonics beyond 22 eV},
  author={Wang, He and Xu, Yiming and Ulonska, Stefan and Robinson, Joseph S and Ranitovic, Predrag and Kaindl, Robert A},
  journal={Nature communications},
  volume={6},
  number={1},
  pages={1--7},
  year={2015},
  publisher={Nature Publishing Group}
}

@article{koll2022phase,
  title={Phase-locking of time-delayed attosecond XUV pulse pairs},
  author={Koll, Lisa-Marie and Maikowski, Laura and Drescher, Lorenz and Vrakking, Marc JJ and Witting, Tobias},
  journal={Optics Express},
  volume={30},
  number={5},
  pages={7082--7095},
  year={2022},
  publisher={Optica Publishing Group}
}

@article{gorlach2020quantum,
  title={The quantum-optical nature of high harmonic generation},
  author={Gorlach, Alexey and Neufeld, Ofer and Rivera, Nicholas and Cohen, Oren and Kaminer, Ido},
  journal={Nature communications},
  volume={11},
  number={1},
  pages={1--11},
  year={2020},
  publisher={Nature Publishing Group}
}

@article{harrison2022increased,
  title={Increased phase precision of spatial light modulators using irrational slopes: application to attosecond metrology},
  author={Harrison, Geoffrey R and Saule, Tobias and Davis, Brandin and Trallero-Herrero, Carlos A},
  journal={Applied Optics},
  volume={61},
  number={30},
  pages={8873--8879},
  year={2022},
  publisher={Optica Publishing Group}
}

@article{lewenstein1994theory,
  title={Theory of high-harmonic generation by low-frequency laser fields},
  author={Lewenstein, Maciej and Balcou, Ph and Ivanov, M Yu and L’huillier, Anne and Corkum, Paul B},
  journal={Physical Review A},
  volume={49},
  number={3},
  pages={2117},
  year={1994},
  publisher={APS}
}

@article{l1993high,
  title={High-order harmonic generation in rare gases with a 1-ps 1053-nm laser},
  author={L’Huillier, Anne and Balcou, Ph},
  journal={Physical Review Letters},
  volume={70},
  number={6},
  pages={774},
  year={1993},
  publisher={APS}
}

@article{krause1992high,
  title={High-order harmonic generation from atoms and ions in the high intensity regime},
  author={Krause, Jeffrey L and Schafer, Kenneth J and Kulander, Kenneth C},
  journal={Physical Review Letters},
  volume={68},
  number={24},
  pages={3535},
  year={1992},
  publisher={APS}
}

@article{davino2021higher,
  title={Higher-order harmonic generation and strong field ionization with Bessel--Gauss beams in a thin jet geometry},
  author={Davino, Michael and Summers, Adam and Saule, Tobias and Tross, Jan and McManus, Edward and Davis, Brandin and Trallero-Herrero, Carlos},
  journal={JOSA B},
  volume={38},
  number={7},
  pages={2194--2200},
  year={2021},
  publisher={Optical Society of America}
}

@article{jansen2016spatially,
  title={Spatially resolved Fourier transform spectroscopy in the extreme ultraviolet},
  author={Jansen, GSM and Rudolf, Denis and Freisem, Lars and Eikema, KSE and Witte, S},
  journal={Optica},
  volume={3},
  number={10},
  pages={1122--1125},
  year={2016},
  publisher={Optical Society of America}
}

@article{mandal2021attosecond,
  title={Attosecond delay lines: Design, characterization and applications},
  author={Mandal, Ankur and Sidhu, Mehra S and Rost, Jan M and Pfeifer, Thomas and Singh, Kamal P},
  journal={The European Physical Journal Special Topics},
  pages={1--19},
  year={2021},
  publisher={Springer}
}

@article{tross2019high,
  title={High harmonic generation spectroscopy via orbital angular momentum},
  author={Tro{\ss}, Jan and Trallero-Herrero, Carlos A},
  journal={The Journal of Chemical Physics},
  volume={151},
  number={8},
  pages={084308},
  year={2019},
  publisher={AIP Publishing LLC}
}

@article{zhu2018investigation,
  title={Investigation of the thermal and optical performance of a spatial light modulator with high average power picosecond laser exposure for materials processing applications},
  author={Zhu, Guangyu and Whitehead, David and Perrie, Walter and Allegre, OJ and Olle, V and Li, Qianliang and Tang, Yue and Dawson, Karl and Jin, Yang and Edwardson, SP and others},
  journal={Journal of Physics D: Applied Physics},
  volume={51},
  number={9},
  pages={095603},
  year={2018},
  publisher={IOP Publishing}
}

@article{kaakkunen2014fast,
  title={Fast micromachining using spatial light modulator and galvanometer scanner with infrared pulsed nanosecond fiber laser},
  author={Kaakkunen, Jarno JJ and Vanttaja, Ilkka and Laakso, Petri},
  journal={Journal of Laser Micro Nanoengineering},
  volume={9},
  number={1},
  pages={37},
  year={2014},
  publisher={Reza Netsu Kako Kenkyukai}
}

@inproceedings{carbajo2018power,
  title={Power handling for LCoS spatial light modulators},
  author={Carbajo, Sergio and Bauchert, Kipp},
  booktitle={Laser Resonators, Microresonators, and Beam Control XX},
  volume={10518},
  pages={282--290},
  year={2018},
  organization={SPIE}
}

@article{orfanos2019attosecond,
  title={Attosecond pulse metrology},
  author={Orfanos, I and Makos, I and Liontos, I and Skantzakis, E and F{\"o}rg, Benjamin and Charalambidis, D and Tzallas, P},
  journal={Apl Photonics},
  volume={4},
  number={8},
  pages={080901},
  year={2019},
  publisher={AIP Publishing LLC}
}

@article{DiPiazza2012,
archivePrefix = {arXiv},
arxivId = {1111.3886},
author = {{Di Piazza}, A. and M{\"{u}}ller, C. and Hatsagortsyan, K. Z. and Keitel, C. H.},
doi = {10.1103/RevModPhys.84.1177},
eprint = {1111.3886},
issn = {00346861},
journal = {Rev. Mod. Phys.},
month = {aug},
number = {3},
pages = {1177--1228},
publisher = {American Physical Society},
title = {{Extremely high-intensity laser interactions with fundamental quantum systems}},
url = {https://journals.aps.org/rmp/abstract/10.1103/RevModPhys.84.1177},
volume = {84},
year = {2012}
}

@article{Ritus1985,
author = {Ritus, V. I.},
doi = {10.1007/BF01120220},
issn = {02702010},
journal = {J. Sov. Laser Res.},
month = {sep},
number = {5},
pages = {497--617},
publisher = {Kluwer Academic Publishers-Plenum Publishers},
title = {{Quantum effects of the interaction of elementary particles with an intense electromagnetic field}},
url = {https://link.springer.com/article/10.1007/BF01120220},
volume = {6},
year = {1985}
}

@article{Wistisen2018,
archivePrefix = {arXiv},
arxivId = {1704.01080},
author = {Wistisen, Tobias N. and {Di Piazza}, Antonino and Knudsen, Helge V. and Uggerh{\o}j, Ulrik I.},
doi = {10.1038/s41467-018-03165-4},
eprint = {1704.01080},
issn = {20411723},
journal = {Nat. Commun.},
month = {feb},
number = {1},
pages = {1--6},
pmid = {29476095},
publisher = {Nature Publishing Group},
title = {{Experimental evidence of quantum radiation reaction in aligned crystals}},
url = {https://www.nature.com/articles/s41467-018-03165-4},
volume = {9},
year = {2018}
}

@article{Poder2018,
archivePrefix = {arXiv},
arxivId = {1709.01861},
author = {Poder, K. and Tamburini, M. and Sarri, G. and {Di Piazza}, A. and Kuschel, S. and Baird, C. D. and Behm, K. and Bohlen, S. and Cole, J. M. and Corvan, D. J. and Duff, M. and Gerstmayr, E. and Keitel, C. H. and Krushelnick, K. and Mangles, S. P.D. and McKenna, P. and Murphy, C. D. and Najmudin, Z. and Ridgers, C. P. and Samarin, G. M. and Symes, D. R. and Thomas, A. G.R. and Warwick, J. and Zepf, M.},
doi = {10.1103/PhysRevX.8.031004},
eprint = {1709.01861},
issn = {21603308},
journal = {Phys. Rev. X},
month = {jul},
number = {3},
pages = {031004},
publisher = {American Physical Society},
title = {{Experimental Signatures of the Quantum Nature of Radiation Reaction in the Field of an Ultraintense Laser}},
url = {https://journals.aps.org/prx/abstract/10.1103/PhysRevX.8.031004},
volume = {8},
year = {2018}
}

@book{born_wolf_2019, 
place={Cambridge}, 
edition={7}, 
title={Principles of Optics: 60th Anniversary Edition}, 
doi={10.1017/9781108769914}, 
publisher={Cambridge University Press}, 
author={Born, Max and Wolf, Emil}, 
year={2019}
}

@book{saleh_teich,
author = {Saleh, Bahaa and Teich, Malvin},
year = {2019},
month = {02},
pages = {},
title = {Fundamentals of Photonics, 3rd Edition},
isbn = {9781119506874}
}

@book{siegman,
author = {Siegman, A. E. },
title = {Lasers / Anthony E. Siegman },
isbn = {0935702115 },
publisher ={ University Science Books Mill Valley, Calif },
year = {1986},
}

@article{Paul2001,
   author = {P. M. Paul and E. S. Toma and P. Breger and G. Mullot and F. Auge and Ph. Balcou and H. G. Muller and P. Agostini},
   doi = {10.1126/science.1059413},
   issn = {0036-8075},
   issue = {5522},
   journal = {Science},
   month = {6},
   pages = {1689-1692},
   title = {Observation of a Train of Attosecond Pulses from High Harmonic Generation},
   volume = {292},
   url = {https://www.science.org/doi/10.1126/science.1059413},
   year = {2001},
}

@article{Kong2017,
    title = {{Controlling the orbital angular momentum of high harmonic vortices}},
    year = {2017},
    journal = {Nature Communications},
    author = {Kong, Fanqi and Zhang, Chunmei and Bouchard, Frédéric and Li, Zhengyan and Brown, Graham G and Ko, Dong Hyuk and Hammond, T J and Arissian, Ladan and Boyd, Robert W and Karimi, Ebrahim and Corkum, P B},
    volume = {8},
    url = {www.nature.com/naturecommunications},
    doi = {10.1038/ncomms14970},
    issn = {20411723},
    pmid = {28378823}
}

@article{Berakdar1998TwoElectrons,
    title = {{Emission of correlated electron pairs following single-photon absorption by solids and surfaces}},
    year = {1998},
    journal = {Physical Review B - Condensed Matter and Materials Physics},
    author = {Berakdar, Jamal},
    number = {15},
    month = {10},
    pages = {9808--9816},
    volume = {58},
    publisher = {American Physical Society},
    url = {https://journals.aps.org/prb/abstract/10.1103/PhysRevB.58.9808},
    doi = {10.1103/PhysRevB.58.9808},
    issn = {1550235X}
}

@article{Mahmood2022TwoElectrons,
    title = {{Distinguishing finite-momentum superconducting pairing states with two-electron photoemission spectroscopy}},
    year = {2022},
    journal = {Physical Review B},
    author = {Mahmood, Fahad and Devereaux, Thomas and Abbamonte, Peter and Morr, Dirk K.},
    number = {6},
    month = {2},
    pages = {064515},
    volume = {105},
    publisher = {American Physical Society},
    url = {https://journals.aps.org/prb/abstract/10.1103/PhysRevB.105.064515},
    doi = {10.1103/PhysRevB.105.064515},
    issn = {24699969},
    arxivId = {2108.04260}
}

@article{Tross2017,
    title = {{N 2 HOMO-1 orbital cross section revealed through high-order-harmonic generation}},
    year = {2017},
    journal = {Physical Review A},
    author = {Tro{\ss}, Jan and Ren, Xiaoming and Makhija, Varun and Mondal, Sudipta and Kumarappan, Vinod and Trallero-Herrero, Carlos A.},
    number = {3},
    month = {3},
    pages = {033419},
    volume = {95},
    publisher = {American Physical Society},
    url = {https://link.aps.org/doi/10.1103/PhysRevA.95.033419},
    doi = {10.1103/PhysRevA.95.033419},
    issn = {2469-9926}
}

@article{Rodnova2020,
    title = {{Generation and control of phase-locked Bessel beams with a persistent non-interfering region}},
    year = {2020},
    journal = {Journal of the Optical Society of America B},
    author = {Rodnova, Zhanna and Saule, Tobias and Sadlon, Richard and McManus, Edward and May, Nicholas and Yu, Xiaoming and Shahbazmohamadi, Sina and Trallero, Carlos},
    number = {11},
    month = {9},
    pages = {3179--3183},
    volume = {37},
    publisher = {The Optical Society},
    url = {https://www.osapublishing.org/viewmedia.cfm?uri=josab-37-11-3179&seq=0&html=true https://www.osapublishing.org/abstract.cfm?uri=josab-37-11-3179 https://www.osapublishing.org/josab/abstract.cfm?uri=josab-37-11-3179},
    doi = {10.1364/josab.400801},
    issn = {0740-3224}
}

@article{Shiner2009,
    title = {{Wavelength scaling of high harmonic generation efficiency.}},
    year = {2009},
    journal = {Physical Review Letters},
    author = {Shiner, A D and Trallero-Herrero, C and Kajumba, N and Bandulet, H-C and Comtois, D and L{\'{e}}gar{\'{e}}, F and Gigu{\`{e}}re, M and Kieffer, J-C and Corkum, P B and Villeneuve, D M},
    number = {7},
    pages = {073902},
    volume = {103},
    publisher = {APS},
    url = {http://link.aps.org/doi/10.1103/PhysRevLett.103.073902},
    institution = {National Research Council of Canada, 100 Sussex Drive, Ottawa, Ontario K1A 0R6, Canada.},
    pmid = {19792645}
}

@article{Shiner2013,
    title = {{High harmonic cutoff energy scaling and laser intensity measurement with a 1.8 {$\mu$}m laser source}},
    year = {2013},
    journal = {Journal of Modern Optics},
    author = {Shiner, A.D. and Trallero-Herrero, C. and Kajumba, N. and Schmidt, B.E. and Bertrand, J.B. and Kim, Kyung Taec and Bandulet, H.-C. and Comtois, D. and Kieffer, J.-C. and Rayner, D.M. and Corkum, P.B. and L{\'{e}}gar{\'{e}}, F. and Villeneuve, D.M.},
    number = {17},
    month = {10},
    pages = {1458--1465},
    volume = {60},
    url = {http://www.tandfonline.com/doi/abs/10.1080/09500340.2013.765067},
    doi = {10.1080/09500340.2013.765067},
    issn = {0950-0340}
}

@article{Greenman2010,
  title = {Implementation of the time-dependent configuration-interaction singles method for atomic strong-field processes},
  author = {Greenman, Loren and Ho, Phay J. and Pabst, Stefan and Kamarchik, Eugene and Mazziotti, David A. and Santra, Robin},
  journal = {Phys. Rev. A},
  volume = {82},
  issue = {2},
  pages = {023406},
  numpages = {12},
  year = {2010},
  month = {Aug},
  publisher = {American Physical Society},
  doi = {10.1103/PhysRevA.82.023406},
  url = {https://link.aps.org/doi/10.1103/PhysRevA.82.023406}
}

@article{Sie2019,
author = {Sie, Edbert J. and Rohwer, Timm and Lee, Changmin and Gedik, Nuh},
doi = {10.1038/s41467-019-11492-3},
issn = {20411723},
journal = {Nat. Commun.},
month = {dec},
number = {1},
pages = {1--11},
pmid = {31388015},
publisher = {Nature Publishing Group},
title = {{Time-resolved XUV ARPES with tunable 24–33 eV laser pulses at 30 meV resolution}},
url = {https://doi.org/10.1038/s41467-019-11492-3},
volume = {10},
year = {2019}
}

@article{Corder2018,
archivePrefix = {arXiv},
arxivId = {1801.08124},
author = {Corder, Christopher and Zhao, Peng and Bakalis, Jin and Li, Xinlong and Kershis, Matthew D. and Muraca, Amanda R. and White, Michael G. and Allison, Thomas K.},
doi = {10.1063/1.5045578},
eprint = {1801.08124},
issn = {23297778},
journal = {Struct. Dyn.},
month = {sep},
number = {5},
pages = {054301},
publisher = {American Crystallographic Association},
title = {{Ultrafast extreme ultraviolet photoemission without space charge}},
url = {http://aca.scitation.org/doi/10.1063/1.5045578},
volume = {5},
year = {2018}
}

@article{Smallwood2012,
author = {Smallwood, Christopher L and Hinton, James P and Jozwiak, Christopher and Zhang, Wentao and Koralek, Jake D and Eisaki, Hiroshi and Lee, Dung-Hai and Orenstein, Joseph and Lanzara, Alessandra},
doi = {10.1126/science.1217423},
issn = {1095-9203},
journal = {Science},
month = {jun},
number = {6085},
pages = {1137},
pmid = {22654053},
title = {{Tracking Cooper pairs in a cuprate superconductor by ultrafast angle-resolved photoemission.}},
url = {http://www.ncbi.nlm.nih.gov/pubmed/22654053},
volume = {336},
year = {2012}
}

@article{Baggesen2011,
doi = {10.1088/0953-4075/44/11/115601},
url = {https://dx.doi.org/10.1088/0953-4075/44/11/115601},
year = {2011},
month = {may},
publisher = {},
volume = {44},
number = {11},
pages = {115601},
author = {Jan Conrad Baggesen and Lars Bojer Madsen},
title = {On the dipole, velocity and acceleration forms in high-order harmonic generation from a single atom or molecule},
journal = {Journal of Physics B: Atomic, Molecular and Optical Physics},
}

@article{Shore1989,
author = { B.W.   Shore  and  K.C.   Kulander },
title = {Generation of Optical Harmonics by Intense Pulses of Laser Radiation},
journal = {Journal of Modern Optics},
volume = {36},
number = {7},
pages = {857-875},
year  = {1989},
publisher = {Taylor & Francis},
optdoi = {10.1080/09500348914550951},
URL = {https://doi.org/10.1080/09500348914550951},
eprint = {https://doi.org/10.1080/09500348914550951}
}

@misc{riley_handbook_2008,
	title = {Handbook of {Frequency} {Stability} {Analysis}},
	url = {https://tsapps.nist.gov/publication/get_pdf.cfm?pub_id=50505},
	language = {en},
	publisher = {Special Publication (NIST SP), National Institute of Standards and Technology, Gaithersburg, MD},
	author = {Riley, William and Howe, David},
	month = jul,
	year = {2008},
}

@article{Gariepy_2014,
  title = {Creating High-Harmonic Beams with Controlled Orbital Angular Momentum},
  author = {Gariepy, Genevieve and Leach, Jonathan and Kim, Kyung Taec and Hammond, T. J. and Frumker, E. and Boyd, Robert W. and Corkum, P. B.},
  journal = {Phys. Rev. Lett.},
  volume = {113},
  issue = {15},
  pages = {153901},
  numpages = {5},
  year = {2014},
  month = {Oct},
  publisher = {American Physical Society},
  doi = {10.1103/PhysRevLett.113.153901},
  url = {https://link.aps.org/doi/10.1103/PhysRevLett.113.153901}
}

@article{watson_high_power_2025,
    title = {High-power femtosecond molecular broadening and the effects of ro-vibrational coupling},
    volume = {12},
    copyright = {© 2025 Optica Publishing Group},
    issn = {2334-2536},
    url = {https://opg.optica.org/optica/abstract.cfm?uri=optica-12-1-5},
    doi = {10.1364/OPTICA.529193},
    language = {EN},
    number = {1},
    urldate = {2025-02-27},
    journal = {Optica},
    author = {Watson, Kevin and Saule, Tobias and Ivanov, Maksym and Schmidt, Bruno E. and Rodnova, Zhanna and Gibson, George and Berrah, Nora and Trallero, Carlos},
    month = jan,
    year = {2025},
    note = {Publisher: Optica Publishing Group},
    pages = {5--10},
}

\end{document}